\documentclass[aps,prx,twocolumn,superscriptaddress, showpacs]{revtex4-2}

\usepackage{graphicx}
\usepackage{dcolumn}
\usepackage{bm}
\usepackage{color}
\usepackage{ulem}
\usepackage[usenames,dvipsnames,svgnames,table]{xcolor}
\definecolor{mypurple}{HTML}{2e3092}
\definecolor{myorange}{HTML}{ff7019}
\definecolor{mygreen}{HTML}{00b708}

\usepackage{amsfonts}       % adds mathematical fonts
\usepackage{amsmath}        % adds mathematical environments
\usepackage{bm}             % adds Greek characters in bold
\usepackage{mathrsfs}       % adds a better calligraphic font for some symbols
\usepackage{setspace}       % manages line spacing
\usepackage{microtype}      % allows not to overflow to the right at the end of the line
\usepackage{enumerate}      % http://ctan.org/pkg/enumerate
\usepackage{soul}           % for highlighting and crossing
\usepackage{xspace}
\usepackage[unicode=true,
            linktocpage,
            linkbordercolor={0.5 0.5 1}, %blue
            citebordercolor={0.5 1 0.5}, %green
            colorlinks = true,
            linkcolor = mypurple,
            citecolor = mygreen,
            urlcolor = mypurple
            ]{hyperref}
\usepackage{changes}
\usepackage{makecell}

\newcommand{\highTc}{high-$T_{\rm c}$\xspace}

\newcommand{\vecck}{{\vec k}}  % Vector symbol for everything but the theoretical Appendix
\begin{document}

%\preprint{}

%-----------------------------------------%
%############### Title ###################%
%-----------------------------------------%

% \title{Superconducting nickelates band structure\\probed by the Seebeck coefficient in the disordered limit}
\title{Electronic band structure of a superconducting nickelate\\probed by the Seebeck coefficient in the disordered limit}
% \title{Probing the band structure of superconducting nickelates\\with the Seebeck coefficient in the disordered limit}
% \title{Seebeck coefficient in a nickelate superconductor:\\ electronic dispersion in the strange metal phase}
% \title{Electronic dispersion in the strange metal phase\\of the superconducting nickelates probed by the Seebeck coefficient}
% \title{Seebeck coefficient of a superconducting quintuple-layer nickelate}

%-----------------------------------------%
%############### Authors #################%
%-----------------------------------------%



\author{G.~Grissonnanche}
\email{gael.grissonnanche@polytechnique.edu}
\affiliation{Laboratory of Atomic and Solid State Physics, Cornell University, Ithaca, NY, USA}
\affiliation{Kavli Institute at Cornell for Nanoscale Science, Ithaca, NY, USA}
\affiliation{Laboratoire des Solides Irradiés, CEA/DRF/lRAMIS, CNRS, École Polytechnique,
Institut Polytechnique de Paris, F-91128 Palaiseau, France}

\author{G.~A.~Pan}
\affiliation{Department of Physics, Harvard University, Cambridge, MA, USA}

\author{H.~LaBollita}
\affiliation{Department of Physics, Arizona State University, Tempe, AZ, USA}

\author{D. Ferenc Segedin}
\affiliation{Department of Physics, Harvard University, Cambridge, MA, USA}

\author{Q. Song}
\affiliation{Department of Physics, Harvard University, Cambridge, MA, USA}

\author{H. Paik}
\affiliation{Platform for the Accelerated Realization, Analysis, and Discovery of Interface Materials, Cornell University, Ithaca, NY, USA}
\affiliation{School of Electrical and Computer Engineering, University of Oklahoma, Norman, OK, USA}

\author{C. M. Brooks}
\affiliation{Department of Physics, Harvard University, Cambridge, MA, USA}

\author{E. Beauchesne-Blanchet}
\affiliation{Laboratoire des Solides Irradiés, CEA/DRF/lRAMIS, CNRS, École Polytechnique,
Institut Polytechnique de Paris, F-91128 Palaiseau, France}

\author{J. L. Santana González}
\affiliation{Laboratoire des Solides Irradiés, CEA/DRF/lRAMIS, CNRS, École Polytechnique,
Institut Polytechnique de Paris, F-91128 Palaiseau, France}

\author{A.~S.~Botana}
\affiliation{Department of Physics, Arizona State University, Tempe, AZ, USA}

\author{J.~A.~Mundy}
\affiliation{Department of Physics, Harvard University, Cambridge, MA, USA}

\author{B.~J.~Ramshaw}
\email{bradramshaw@cornell.edu}
\affiliation{Laboratory of Atomic and Solid State Physics, Cornell University, Ithaca, NY, USA}
\affiliation{Canadian Institute for Advanced Research, Toronto, Ontario, Canada}


\date{\today}

% %-----------------------------------------%
% %############### Abstract ################%
% %-----------------------------------------%

\begin{abstract}

Superconducting nickelates are a new family of strongly correlated electron materials with a phase diagram closely resembling that of superconducting cuprates. While analogy with the cuprates is natural, very little is known about the metallic state of the nickelates, making these comparisons difficult. We probe the electronic dispersion of thin-film superconducting 5-layer ($n=5$) and metallic 3-layer ($n=3$) nickelates by measuring the Seebeck coefficient, $S$. We find a temperature-independent and negative $S/T$ for both $n=5$ and $n=3$ nickelates. These results are in stark contrast to the strongly temperature-dependent $S/T$ measured at similar electron filling in the cuprate La$_{1.36}$Nd$_{0.4}$Sr$_{0.24}$CuO$_4$. The electronic structure calculated from density functional theory can reproduce the temperature dependence, sign, and amplitude of $S/T$ in the nickelates using Boltzmann transport theory. This demonstrates that the electronic structure obtained from first-principles calculations provides a reliable description of the Fermiology of superconducting nickelates, and suggests that, despite indications of strong electronic correlations, there are well-defined quasiparticles in the metallic state. Finally, we explain the differences in the Seebeck coefficient between nickelates and cuprates as originating in strong dissimilarities in impurity concentrations. Our study demonstrates that the high elastic scattering limit of the Seebeck coefficient reflects only the underlying band structure of a metal, analogous to the high magnetic field limit of the Hall coefficient. This opens a new avenue for Seebeck measurements to probe the electronic band structures of relatively disordered quantum materials.

%Beyond establishing a baseline understanding of how the electronic structure relates to transport coefficients in these new materials, this work demonstrates the power of the semi-classical approach to quantitatively describe transport measurements, even in proximity to the strange-metallic state.

\end{abstract}

%-----------------------------------------%
%############### PACS ####################%
%-----------------------------------------%

\pacs{74.72.Gh, 74.25.Dw, 74.25.F-}

% 74.72.Gh  Hole-doped cuprate superconductors
% 74.25.Dw  Phase diagrams superconductivity
% 74.25.F-  Transport properties

%-----------------------------------------%
%############### Keywords ################%
%-----------------------------------------%

%\keywords{}

%-----------------------------------------%
%############### Make Title ##############%
%-----------------------------------------%

\maketitle

%-----------------------------------------%
%############### MAIN ####################%
%-----------------------------------------%

% References should be done using the \cite,We synthesized t, \ref, and \label commands

%>>>>>>>>>>>>>>>>>>>>>>>>>>>>>>>>>>>>>>>>>>>>>>>>>>>>>>>>>>>>>>>>>>>>>>>>>>>>>>>>>>>>>>>>>>>>>>>>>>
\section{INTRODUCTION}

Unconventional superconductivity remains one of the most active and challenging subfields of strongly correlated electron research, with cuprates posing some of the toughest experimental and theoretical challenges over the past three decades \cite{Keimer2015From}. The origin of \highTc superconductivity in the cuprates remains a mystery in part due to the complex interplay of several competing states and relatively strong disorder. One approach to understanding the physics of \highTc is to replace copper entirely, for example with ruthenium or nickel, while maintaining the same square-lattice, transition metal oxide motif. Sr$_2$RuO$_4$ is a success of this approach \cite{maeno_superconductivity_1994}, but it does not share the complex phase diagram of the cuprates.

The recent discovery of superconductivity in strontium-doped NdNiO$_{2}$ \cite{Li2019Superconductivity,Danfeng2020,Zeng2020} and stoichiometric Nd$_6$Ni$_5$O$_{12}$ \cite{pan_superconductivity_2022} presents an opportunity to explore the key ingredients for unconventional superconductivity by contrasting the physical properties of the nickelates with the cuprates. The nickelates contain cuprate-like NiO$_2$ planes, and the family we study here is Nd$_{\rm n+1}$Ni$_n$O$_{\rm 2n+2}$, where $n$ indicates the number of NiO$_2$ planes per unit cell \cite{Poltavets2007crystal, Poltavets2009electronic, Greenblatt2010bulk, Labollita2021electronic, Labollita2022manybody}. While nickel in the $n = \infty$ member of the series---NdNiO$_{2}$---has the same nominal $3d^9$ electronic configuration as copper does in the cuprates, the finite-$n$ members have the nominal configuration of $3d^{9-\delta}$, where $\delta = 1/n$. This offers a mechanism for exploring the hole-doped phase diagram without introducing cation disorder.

\begin{figure}[t]
\centering
\includegraphics[width=0.435\textwidth]{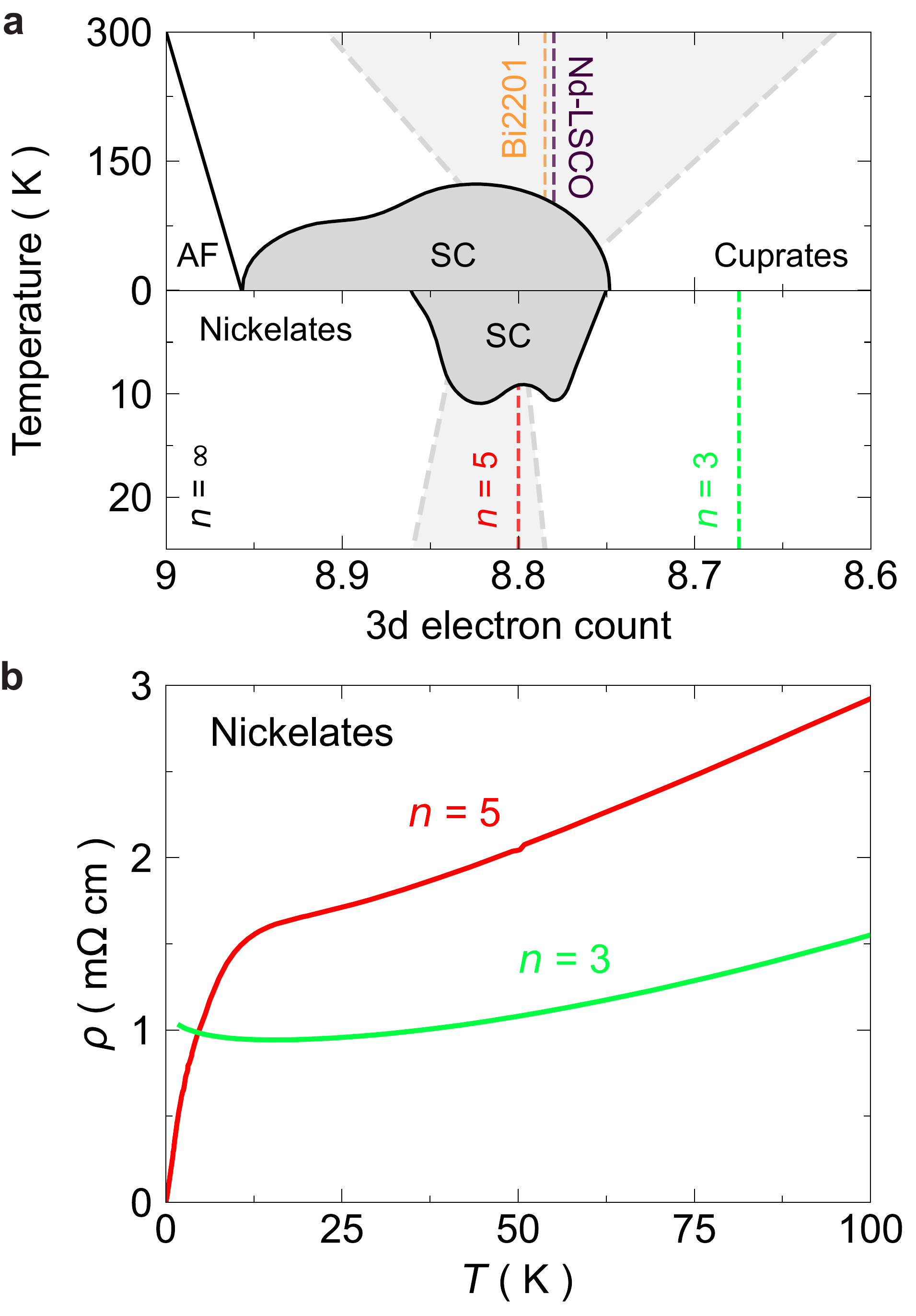}
\caption{
(\textbf{a}) Schematic temperature versus 3\textit{d} electron count phase diagrams of cuprates (top) and nickelates (bottom). Different phases are displayed: superconducting phase (SC, dark grey), strange metal (light grey delimited by dashed lines \cite{Cooper2009Anomalous,lee_linear_character_2023}), antiferromagnetism (AF). The location in these phase diagrams of the studied sample are represented by vertical dashed lines Nd$_6$Ni$_5$O$_{12}$ (Ni$^{1.2+}$: d$^{8.8}$, red) and Nd$_4$Ni$_3$O$_{8}$  (Ni$^{1.33+}$: d$^{8.67}$, green), indicated as $n=5$ and $n=3$, respectively, Nd-LSCO $p=0.24$ (purple), Bi2201 $p=0.23$ (orange).
(\textbf{b}) In-plane resistivity vs $T$ at $B = 0$~T of Nd$_6$Ni$_5$O$_{12}$ ($n = 5$ nickelate, red), Nd$_4$Ni$_3$O$_8$ ($n=3$ nickelate, green) as measured by Pan \textit{et al.}~\cite{pan_superconductivity_2022}.
}
\label{fig:phase_diagram}
\end{figure}

Superconducting nickelates exhibit many similarities with the cuprates. These include a phase diagram with a superconducting dome maximized around similar $3d^{8.8}$ electron concentrations, evidence for a nodal superconducting gap~\cite{botana_similarities_2020}, magnetism\cite{lu_magnetic_2021,fowlie_intrinsic_2022}, charge density waves \cite{rossi_broken_2022,tam_charge_2022}, and even a strange metal phase~\cite{lee_linear_character_2023} (Fig.~\ref{fig:phase_diagram}a). Conspicuously absent from this list are experimental comparisons of the electronic structure. To understand which aspects of the electronic dispersion are favorable for unconventional superconductivity, one must first understand how electrons interact in the normal metallic state.

The central difficulty is that most of the experimental techniques used to study electronic structures are incompatible with current superconducting nickelate samples. There have been attempts to measure the angle-integrated density of states~\cite{chen_electronic_2022}, and there are recent angle-resolved photoemission spectroscopy (ARPES) measurements on non-superconducting, single crystal nickelates~\cite{li_electronic_2022}, but ARPES remains out of reach for superconducting nickelate films due to surface quality issues. Similarly, quantum oscillations require metals with a defect density lower than what is currently available in even the cleanest films. This calls for the use of other techniques that are sensitive to the electronic structure and that are compatible with higher levels of elastic scattering from defects and with thin films.

\begin{figure}[t]
\centering
\includegraphics[width=0.47\textwidth]{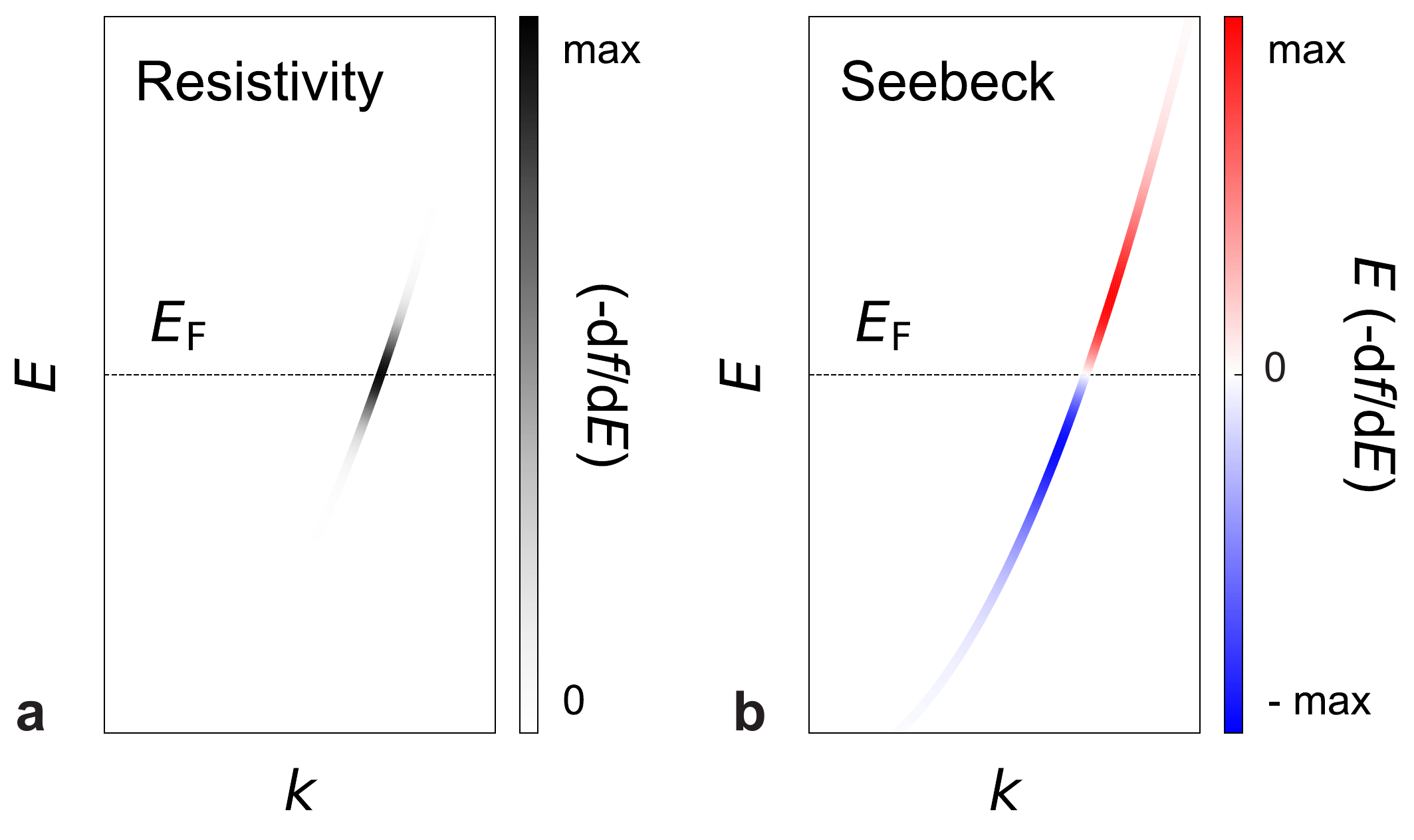}
\caption{
Sketch of a band dispersion, highlighting the electronic states that contribute the most to (\textbf{a}) resistivity and (\textbf{b}) the Seebeck coefficient, as indicated by the color gradients. The states are selected by the weighting factors $(-\frac{df}{dE})$ and $-E(\frac{df}{dE})$ from the equations \ref{eq:sigma} for the resistivity and \ref{eq:alpha} for the thermoelectric coefficient, respectively, at a given temperature $T$. The states that contribute most to the resistivity (Seebeck coefficient) are located at the Fermi level (on either side of the Fermi level). In the case of the Seebeck coefficient, the contributions of states above the Fermi level are subtracted from the contributions of states below the Fermi level----hence the Seebeck coefficient is a measure of the particle-hole asymmetry.
}
\label{fig:band_dispersion_sketch}
\end{figure}

Thermoelectricity---as measured by the Seebeck coefficient $S$---provides an alternative to probe the electronic band structure of a material.
Unlike electrical transport, which is only sensitive to the electronic states in the immediate vicinity of the Fermi energy ($E_F$) (Fig.~\ref{fig:band_dispersion_sketch}a), the Seebeck effect is sensitive to details of the electronic dispersion away from $E_F$.
Specifically, the Seebeck coefficient reflects the asymmetry of the dispersion above and below $E_F$---it probes the asymmetry between occupied and unoccupied states (Fig.~\ref{fig:band_dispersion_sketch}b), also called particle-hole asymmetry or energy asymmetry~\cite{Gourgout2022Seebeck,Kondo2005Contribution}.
In general, the Seebeck coefficient is defined by both the band structure and the energy dependence of the scattering rate. However, we will demonstrate that this coefficient is only determined by the band structure in the disordered limit, which is analogous to how the Hall coefficient becomes independent of scattering rate in the high-field limit. As the high-field limit is usually inaccessible in most metals, this makes the Seebeck effect a new powerful probe of the electronic dispersion of relatively disordered materials.
%

%#################### Figure 1 ####################%
%##################################################%

To investigate the electronic structure of the nickelates, we measured the Seebeck coefficient of a superconducting 5-layer nickelate Nd$_6$Ni$_5$O$_{12}$ ($n=5$ nickelate) with a transition onsetting at $T_{\rm c} \approx 10$~K (Fig. \ref{fig:phase_diagram}b)---as well as a more-overdoped, non-superconducting, 3-layer nickelate Nd$_4$Ni$_3$O$_{8}$ ($n=3$ nickelate) for comparison (Fig. \ref{fig:phase_diagram}b). % The 3-layer nickelates have been shown to be the closest cuprate analog to date \cite{Zhang2016stacked, Zhang2017large, Sarkar2011electronic}.
We find that both the $n=5$ and $n=3$ nickelates share a similar temperature independent, negative $S/T$. We show the electronic dispersion obtained from density functional theory (DFT) accounts for both the magnitude and sign of the temperature-independent Seebeck coefficient for the two compounds when calculated in the disordered limit.

To justify the disordered limit, we compare the nickelate data to previous measurements of the Seebeck coefficient in hole doped cuprates with a similar electron count to the $n=5$ nickelate. First, we compare with measurements performed on a single crystal of La$_{1.36}$Nd$_{0.4}$Sr$_{0.24}$CuO$_4$ (Nd-LSCO $p=0.24$) \cite{Gourgout2022Seebeck} with a Seebeck coefficient that is positive and qualitatively different from that of the nickelates. Second, we compare with measurements performed on a single crystal of (Bi,Pb)$_2$(Sr,La)$_2$CuO$_{6+\delta}$ (Bi2201 $p=0.23$) \cite{Kondo2005Contribution,berben_2022a} with an almost identical Seebeck coefficient to the nickelates. Despite their disparities, we show that the differences in Seebeck coefficients between nickelates and cuprates come from strong dissimilarities in impurity concentrations, and not necessarily from fundamental differences in the nature of the metallic state. Despite the presence of strong electronic correlations, the success of DFT and semi-classical transport calculations in our study provides evidence of well-defined quasiparticles responsible for charge and heat transport in both nickelates and cuprates.

%#################### Figure 2 ####################%
%##################################################%

%>>>>>>>>>>>>>>>>>>>>>>>>>>>>>>>>>>>>>>>>>>>>>>>>>>>>>>>>>>>>>>>>>>>>>>>>>>>>>>>>>>>>>>>>>>>>>>>>>>

\section{METHODS}

\textbf{Samples}.
%The layered nickelate samples were prepared in thin-film form with molecular beam epitaxy, followed by a subsequent \textit{ex situ}, CaH$_2$-assisted, topotactic reduction process.
%
The perovskite-like parent Nd$_{\rm n+1}$Ni$_n$O$_{\rm 3n+1}$ films ($n=5$ and $n=3$) were synthesized by molecular beam epitaxy on (110)-orientated NdGaO$_3$. The growth process used distilled ozone, substrate temperatures of $\sim$650-690 $^{\circ}$C, and the NdNiO$_3$ calibration procedure described in Ref.~\cite{Pan2022Synthesis}.
This synthesis was followed by a reduction process contained in a sealed glass ampoule, optimized with a process at $\sim290^\circ$C lasting three hours in order to reach the square-planar Nd$_{\rm n+1}$Ni$_n$O$_{\rm 2n+2}$ phases (this process is similar to the procedure in Ref.~\cite{pan_superconductivity_2022}).
Using an electron-beam evaporator, contacts consisting of a 10 nm chromium sticking layer and 150~nm of gold were deposited in a Hall bar geometry such that the applied thermal gradient and measured Seebeck voltage were along the [001]-direction of the substrate.

The substrate material NdGaO$_3$ has a high thermal conductivity that increases 30-fold between room temperature and $\sim$30~K~\cite{Schnelle_2001}, weakening the applied thermal gradient along the nickelate film.
To mitigate this effect, we polished the NdGaO$_3$ substrate to reduce its thickness from $500$ microns down to $\sim$100 - 150 microns using diamond lapping film. This served to increase the thermal gradient that generates the Seebeck voltage, which allowed us to measure the Seebeck effect down to $\sim$60~K, below which the thermal gradient becomes too small and the experiment cannot be performed reliably.
This process necessarily involves a brief heat exposure during sample mounting.
We minimized the degradation risk to the sample~\cite{ding_stability_2022} by using low temperature crystal wax and mounting in an argon glove box; resistivity measurements taken before and after polishing showed no substantial changes.

\textbf{Measurements}.
We measured the Seebeck coefficient using an AC technique used previously for cuprates~\cite{Gourgout2022Seebeck}.
An AC thermal excitation is generated by passing an electric current at frequency $\omega \sim 0.1$~Hz through a 5 k$\Omega$ strain gauge used as a heater to generate a thermal gradient in the sample. While the heat is carried primarily by the substrate, this also generates a thermal gradient $\Delta T_{\rm AC}$ along the film. We detect this AC thermal gradient at frequency $2\omega$, as well as the absolute temperature shift, using two type E thermocouples.
An AC Seebeck voltage, $\Delta V_{\rm AC}$, is also generated at a frequency $2\omega$ in response to the thermal gradient. We measure this voltage with phosphor-bronze wires attached to the same contacts where the thermocouples measure $\Delta T_{\rm AC}$: this eliminates uncertainties associated with the geometric factor.

The thermocouple and Seebeck voltages were amplified using EM Electronics A10 preamplifiers and detected using a MCL1-540 Synktek lock-in amplifier at the thermal excitation frequency $2\omega$. The Seebeck coefficient is then given by $S = - \Delta V_{\rm AC} / \Delta T_{\rm AC}$. The frequency $\omega$ was adjusted so that the thermoelectric voltage and the thermal gradient remained in phase.

\textbf{Band structure calculations}. The paramagnetic electronic structure of the $n = 5$ and $n = 3$ layered nickelates was calculated using density functional theory (DFT) combined with the projector augmented wave method, as implemented in the Vienna \textit{ab-initio} simulation package \cite{Kresse1996}. We used a pseudopotential that treats the Nd $4f$ electrons as core electrons. The in-plane lattice parameters were set to match the NdGaO$_3$ substrate, and we optimized the out-of-plane lattice parameter. See Appendix~\ref{app:dft_details} for more details on the band structure calculations.

\textbf{Boltzmann transport}. We fit a tight-binding model (Tables \ref{tab:tight-binding_5} and \ref{tab:tight-binding_3}) to the DFT band structure calculated for the nickelates (Fig.~\ref{fig:dft_data}). We combined the tight-binding model and Boltzmann transport theory to calculate the Seebeck coefficient. We applied the same algorithm that was used successfully in the cuprates \cite{Grissonnanche2021LinearIn,Fang2022Fermi,Gourgout2022Seebeck,Ataei2022Electrons} to numerically evaluate the Seebeck coefficient for the nickelates.

%#################### Figure 3 ####################%
\begin{figure}[t]
\centering
\includegraphics[width=0.435\textwidth]{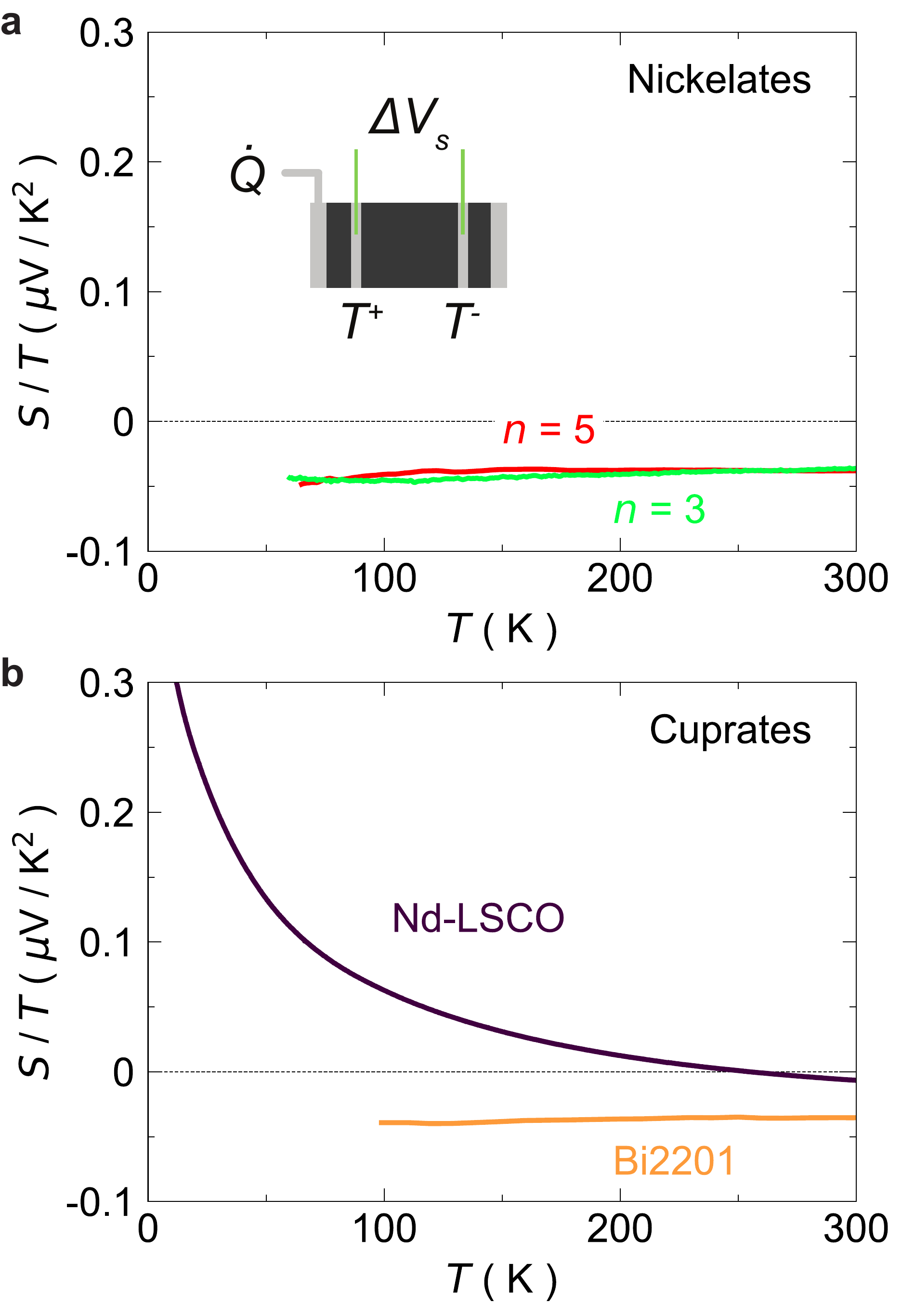}
\caption{
In-plane Seebeck coefficient plotted as $S/T$ vs $T$ of
(\textbf{a}) nickelates $n=5$ (red) and $n=3$ (green) at $B = 0$~T;
(\textbf{b}) cuprates Nd-LSCO $p=0.24$ (purple) measured by Gourgout \textit{et al.}~\cite{Gourgout2022Seebeck} at $B = 16$~T (a field large enough to suppress $T_{\rm c} = 11$~K) and Bi2201 $p=0.23$ measured by Kondo~\textit{et al.} at $B=0$~T (orange). The Seebeck coefficient always goes to zero at zero temperature; we plot $S/T$ to facilitate easier comparison between different materials. The inset in panel (a) shows a schematic of the experimental setup. A heater attached to one end of the sample applies a heat current $\dot{Q}$. The heat current sets up a thermal gradient $\Delta T = T_+ - T_-$, where $T_+$ ($T_-$) is the hot (cold) temperature. A voltage drop $\Delta V_s$ develops in response to $\Delta T$. The Seebeck coefficient is given by $S = - \Delta V_s / \Delta T$.
}
\label{fig:resistivity_seebeck_data}
\end{figure}
%##################################################%

%>>>>>>>>>>>>>>>>>>>>>>>>>>>>>>>>>>>>>>>>>>>>>>>>>>>>>>>>>>>>>>>>>>>>>>>>>>>>>>>>>>>>>>>>>>>>>>>>>>
\section{RESULTS}

\textbf{Seebeck coefficient}.
Fig.~\ref{fig:resistivity_seebeck_data}b shows the in-plane Seebeck coefficient of both the $n=5$ and $n=3$ samples. Both samples show an $S/T$ that is similar in magnitude, negative, and independent of temperature. We reproduced the Seebeck coefficient of the $n=5$ layer nickelate on a second sample (Appendix~\ref{app:sample_comparison}), and the measured $S/T$ of the $n=3$ sample is similar to what was measured previously on the 3-layer nickelate La$_4$Ni$_3$O$_8$ above its metal-to-insulator transition at 105~K~\cite{cheng2012pressure}. 

The Seebeck coefficients of both nickelate samples are also comparable in magnitude and sign to that of the overdoped cuprate Bi2201 $p=0.23$~\cite{Kondo2005Contribution}. All of these measurements contrast with the optimally-doped cuprate Nd-LSCO $p=0.24$~\cite{Gourgout2022Seebeck}, whose Seebeck coefficient is strongly temperature dependent and changes sign near room temperature (Fig.~\ref{fig:resistivity_seebeck_data}b). Both cuprates have a similar electron count to the $n=5$ nickelate.

The Seebeck coefficient in overdoped cuprates has been a puzzle for decades, with most cuprates showing a positive Seebeck coefficient similar to Nd-LSCO at low temperature, but Bi2201 showing a negative Seebeck coefficient. Our analysis is able to account for the differences in sign between Bi2201 and Nd-LSCO and explain the temperature dependence of Bi2201 for the first time, presenting a unified picture of the Seebeck coefficient across nickelates and overdoped cuprates.

%#################### Figure 4 ####################%
\begin{figure*}[t]
\centering
\includegraphics[width=1\textwidth]{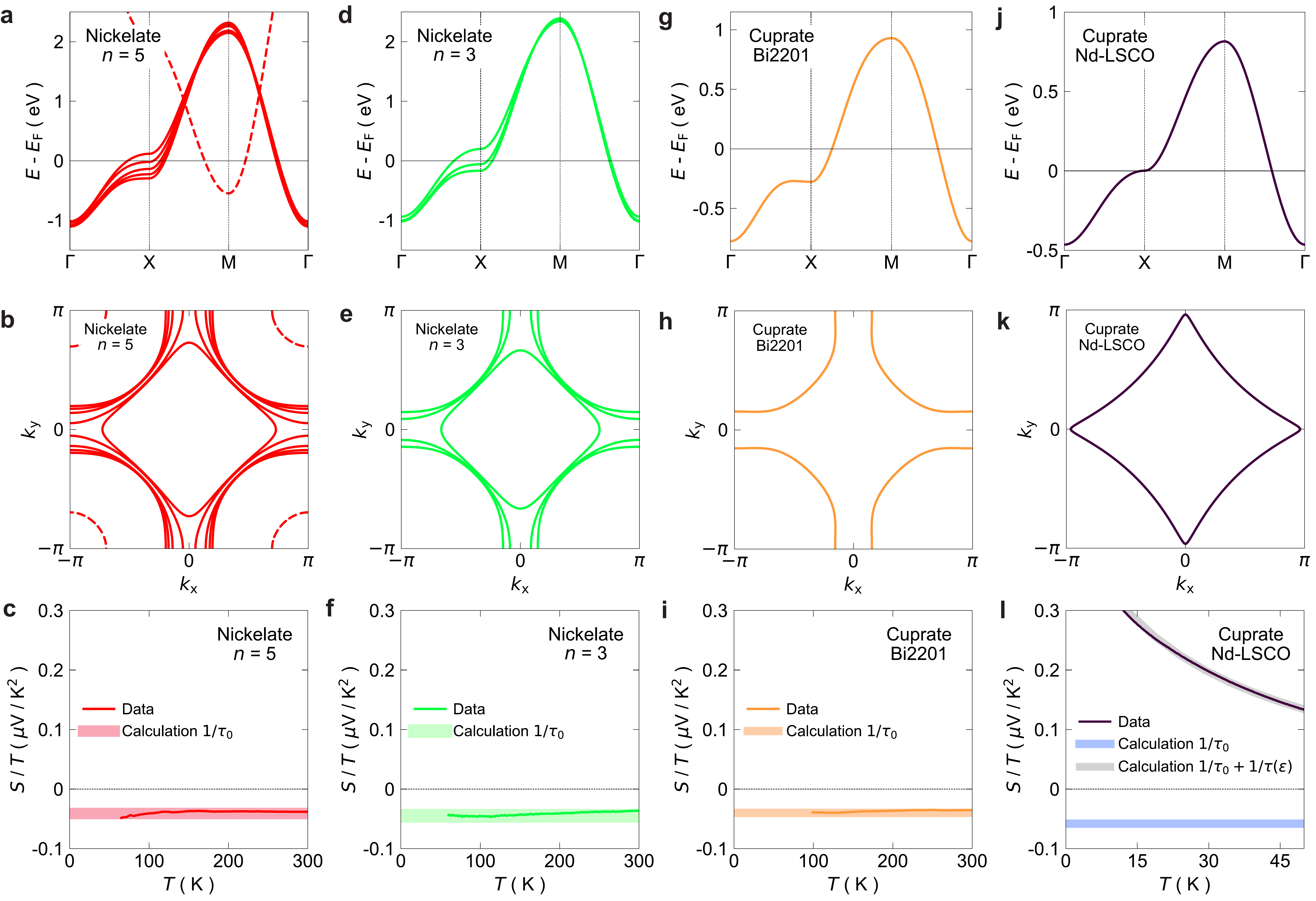}
\caption{Electronic band structures (top), Fermi surfaces (middle) and calculated Seebeck coefficient (bottom) plotted as $S/T$ vs $T$ for (\textbf{a}, \textbf{b}, \textbf{c}) $n = 5$ nickelate; (\textbf{d}, \textbf{e}, \textbf{f}) $n=3$ nickelate; (\textbf{g}, \textbf{h}, \textbf{i}) cuprate Bi2201 $p=0.23$; (\textbf{j}, \textbf{k}, \textbf{l}) cuprate Nd-LSCO $p=0.24$. The DFT calculated bands for the $n=5$ nickelate includes 5 sheets of Ni $d_{\rm x^2-y^2}$ character (1 electron-like, 4 hole like) centered around (0,0) and one Nd-$d$ band (dashed) centered at the zone corner. The DFT calculated bands for the $n=3$ nickelate include 3 sheets of Ni $d_{\rm x^2-y^2}$ (1 electron-like, 2 hole-like). The band for Nd-LSCO is obtained from angle dependent magnetoresistance~\cite{Grissonnanche2021LinearIn}, in agreement with ARPES~\cite{Horio2018ThreeDimensional} measurements. The band for Bi2201 is obtained from ARPES measurements~\cite{Kondo2005Contribution}. The bottom panels compare the measured Seebeck coefficient to the one calculated from the material's bands using Boltzmann transport and a constant elastic scattering rate $1/\tau_0$. Panel (l) for Nd-LSCO $p=0.24$ also displays calculations with a total scattering rate that includes an elastic part $1/\tau_{\rm 0}$ and a particle-hole asymmetric part, $1/\tau(\epsilon)$, which was showed to reproduce the experimental positive Seebeck coefficient by Gourgout \textit{et al.}~\cite{Gourgout2022Seebeck} .
}
\label{fig:dft_seebeck}
\end{figure*}
%##################################################%

\textbf{Boltzmann calculations}. We performed Boltzmann transport calculations to interpret the temperature dependence and the negative sign of $S/T$ in both the $n=5$ and $n=3$ nickelates (see Appendix~\ref{app:boltzmann_calc} for more details). For a free-electron model (i.e. a circular Fermi surface), the sign of the Seebeck coefficient reflects the sign of the charge carriers---hole (positive) or electron (negative)---which is similar to the Hall coefficient. For a real material, the Seebeck coefficient is sensitive to the particle-hole asymmetry of the electronic dispersion (Fig.~\ref{fig:band_dispersion_sketch}b), as well as to the particle-hole asymmetry of the scattering rate, and the resulting Seebeck coefficient can be of either sign.

% add the alpha, sigma formulas in zero magnetic field
% add the graphs for which that states matter for
% explain that DFT gives v_i = depsilon/dk, and that tau_0 if energy independent

To perform Boltzmann transport calculations of the Seebeck coefficient, we require the electronic band dispersions for each material. For $n=5$ and $n=3$ nickelates, we fit a tight-binding model $E(\bf{k})$ to the calculated DFT band structure \cite{Labollita2021electronic, Labollita2022manybody, pan_superconductivity_2022}. For both materials, a single $d_{x^{2}-y^{2}}$ band per NiO$_{2}$ layer crosses $E_{\rm F}$ (Fig.~\ref{fig:dft_seebeck}a and \ref{fig:dft_seebeck}a). For the $n=5$ compound, one additional band of Nd character crosses $E_{\rm F}$ while for the $n=3$ material the Nd bands are well above the Fermi level (see Appendix \ref{app:dft_details} for more details). For the cuprates, we used the tight-binding models obtained from fitting angle dependent magnetoresistance and ARPES for Nd-LSCO~\cite{Grissonnanche2021LinearIn,Horio2018ThreeDimensional}, and ARPES for Bi2201~\cite{Kondo2005Contribution} (Fig.~\ref{fig:dft_seebeck}g and \ref{fig:dft_seebeck}g). The tight-binding model $E(\bf{k})$ provides the velocities $\bf{v} = \frac{1}{\hbar}\nabla_{\rm \bf{k}} E(\bf{k})$ that serves to calculate the Seebeck coefficient.

We obtain excellent agreement between the calculated and measured $S/T$ for the nickelates by using the DFT band dispersions and a constant (energy and temperature independent) scattering rate, $1/\tau_{\rm 0}$ (Fig.~\ref{fig:dft_seebeck}c, \ref{fig:dft_seebeck}f). We will justify this choice of scattering rate below.

%Note that, in the limit where the scattering rate is predominately energy-independent, the Seebeck coefficient becomes independent of scattering because it is the ratio two quantities that are inversely proportional to the scattering rate: the Peltier coefficient $\alpha \propto \tau_{\rm 0}$ and the electric conductivity $\sigma \propto \tau_{\rm 0}$.
%From the temperature dependence of the resistivity, we know that the \textit{total} scattering rate of the nickelates must be temperature (and thus energy) dependent (\autoref{fig:Fig_S_rho_fits}). The temperature-independence of $S/T$ therefore requires that the energy-dependent scattering be a small correction to a larger elastic scattering rate, which is why it drops out of the calculation of $S/T$ (add figure to SI).
%
For the cuprates, the calculations with a constant scattering rate predicts also exactly the right magnitude and sign for Bi2201; while the constant scattering rate calculation for Nd-LSCO gets both the sign and the temperature dependence incorrect.
To obtain agreement between the Nd-LSCO data and Boltzmann transport calculations, an inelastic, particle-hole asymmetric scattering rate must be invoked (Fig.~\ref{fig:dft_seebeck}l from \citet{Gourgout2022Seebeck}).
In this case, the scattering rate is not only energy dependent in Nd-LSCO $p=0.24$ (denoted $1/\tau(\epsilon)$), but it is also linear-in-energy, with a different slope above and below the Fermi energy (see Appendix~\ref{app:boltzmann_calc} and \citet{Gourgout2022Seebeck} for more details).
%
%As the Seebeck coefficient reflects the material's particle-hole asymmetry---arising from either the band dispersion or from the electron scattering rate---an anisotropic and inelastic scattering rate $1/\tau(\epsilon)$ was the key ingredient to explain the sign and temperature dependence of the Seebeck coefficient for Nd-LSCO $p=0.24$.
%
%In the case of the nickelates, because we do not need to invoke an energy-dependent scattering rate to reproduce the data, it implies that the particle-hole asymmetry of the electronic dispersion calculated from DFT is sufficient to capture the experimental value of the Seebeck coefficient.

% >>>>>>>>>>>>>>>>>>>>>>>>>>>>>>>>>>>>>>>>>>>>>>>>>>>>>>>>>>>>>>>>>>>>>>>>>>>>>>>>>>>>>>>>>>>>>>>>>
\section{DISCUSSION}

%\onecolumngrid % important to put a space below that statement
%\twocolumngrid

\textbf{Effect of impurity scattering}. The stark difference in $S/T$ between the nickelates and cuprates is somewhat surprising given the similarity of their electronic structures. These compounds have predominantly 3$d^9$ bands crossing the Fermi energy, and the curvatures of the Fermi surfaces are not all that different---the single Fermi surface in Nd-LSCO essentially interpolates between the hole and electron-like Fermi surfaces found in the multi-layer nickelates (Fig.~\ref{fig:dft_seebeck}a, b, c). Given that the band structure is largely temperature-independent, the disparities in $S/T$ between the two families must originate in a difference in the scattering rate.

To understand this difference, we examine the relative amounts of disorder in the cuprate and nickelate samples by comparing the residual resistivities, $\rho_0$. For the $n=5$ and $n=3$ nickelates, $\rho_0 = 1450$~$\mu\Omega$ cm and $920$~$\mu\Omega$ respectively as measured by Pan \textit{et al.} \cite{pan_superconductivity_2022}, which is significantly larger than the $\rho_0 = 23$ $\mu\Omega$ cm of Nd-LSCO $p=0.24$~\cite{Daou2009Linear} and $\rho_0 \approx 120$ $\mu\Omega$ cm of Bi2201 $p=0.23$~\cite{kondo_resistivity2006,Ayres2021Incoherent,berben_2022a}. To quantify the disorder, we use the same Boltzmann transport framework we use to calculate $S/T$ to fit the elastic scattering rate $1/\tau_0$ to $\rho_0$ for each material. We find that $1/\tau_0$ is approximately $350$ times higher for the $n = 5$ nickelate, and $180$ times for $n=3$, compared to Nd-LSCO $p=0.24$ ($1/\tau_{\rm 0} = 10$~ps$^{-1}$).

In the disordered limit, the scattering rate is predominately energy-independent (elastic), $1/\tau_0 \gg 1/\tau(\epsilon)$, the Seebeck coefficient becomes independent of scattering because it is the ratio of two quantities that are proportional to the scattering time: $S =  \alpha/\sigma$, with the Peltier coefficient $\alpha \propto \tau_{\rm 0}$ and the electrical conductivity $\sigma \propto \tau_{\rm 0}$ (more details are given in Appendix \ref{app:boltzmann_calc}). The high $\rho_0$ and the measured temperature-independent $S/T$ suggest that elastic scattering is indeed dominant in the nickelates, whereas we know the inelastic scattering plays a dominant role in the physics of Nd-LSCO~\cite{Gourgout2022Seebeck}. The cuprate Bi2201, with a significantly higher level of disorder than Nd-LSCO, lies also in the disordered limit like the nickelates, which is confirmed by its temperature independent $S/T$.

\begin{figure}[t!]
\centering
\includegraphics[width=0.435\textwidth]{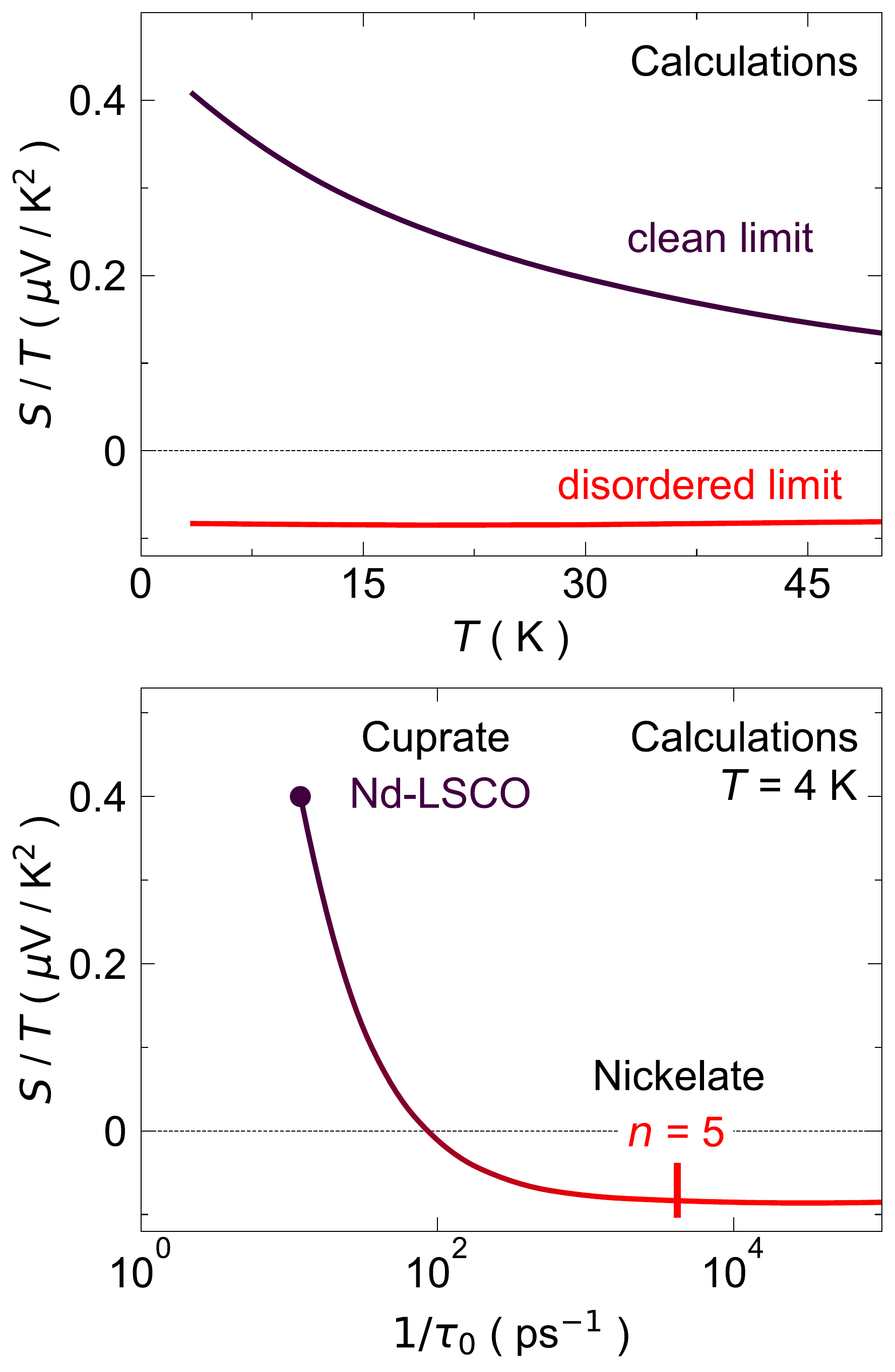}
\caption{
Calculated Seebeck coefficient of Nd-LSCO $p=0.24$, plotted as $S/T$ (\textbf{a}) as function of temperature and (\textbf{b}) as a function of elastic scattering $1/\tau_{\rm 0}$ at $T=4$~K. For both panels, the total scattering rate is given by the $1/\tau_{\rm 0} + 1/\tau(\epsilon)$. In the top panel, the limit of dominant inelastic scattering rate represents $1/\tau_{\rm 0} \sim 1/\tau(\epsilon)$  (purple) and of dominant elastic scattering $1/\tau_{\rm 0} \gg 1/\tau(\epsilon)$) (red). In the ``clean'' limit (purple), we use the elastic scattering rate extracted from ADMR~\cite{Grissonnanche2021LinearIn} on Nd-LSCO $p=0.24$, which gives $1/\tau_{\rm 0}=10$~ps$^{-1}$. In the disordered limit (red), we use the elastic scattering rate extracted from the residual resistivity of the $n=5$ nickelate, which is $1/\tau_{\rm 0}=3500$~ps$^{-1}$. In panel (b), the grey line is obtained from the Nd-LSCO $p=0.24$ calculations at $T=4$~K -- that include $1/\tau_{\rm 0} + 1/\tau(\epsilon)$, by changing the values of $\tau_{\rm 0}$. The $n=5$ nickelate is placed on that curve in regard to its elastic scattering rate value to illustrate its disordered limit.
}
\label{fig:impurity}
\end{figure}

To further illustrate this argument, we recalculate the Seebeck coefficient for Nd-LSCO $p=0.24$ with the total scattering rate $1/\tau_{\rm 0}+1/\tau(\epsilon)$ from \citet{Gourgout2022Seebeck} where we increase the relative amount of elastic scattering $1/\tau_{\rm 0}$ compared to the amount of inelastic scattering $1/\tau(\epsilon)$, going from the clean limit of Nd-LSCO to the disordered limit of the nickelates.
Increasing $1/\tau_{\rm 0}$ from  $10$~ps$^{-1}$ to $3500$~ps$^{-1}$ while holding $1/\tau(\epsilon)$ fixed, the calculated $S/T$ drops to a temperature-independent, negative value---very similar to what we measured in the nickelates (Fig.~\ref{fig:impurity}a) and what was measured for Bi2201.
This confirms that the nickelate films and Bi2201 are dominated by elastic scattering and, in this limit, $S/T$ directly reflects the properties of the electronic bands rather than the energy dependence of the scattering rate.
This proves the effectiveness of our approach to probe the electronic band dispersion of the nickelates using the Seebeck coefficient in the disordered limit. Note that the elastic scattering rate in the infinite layer nickelates is about 10 times smaller~\cite{lee_linear_character_2023} than in the $n=5$ nickelate.
However, Fig.~\ref{fig:impurity}b shows that the infinite layer nickelates are still in the limit where the elastic scattering rate dominates over the energy-asymmetric scattering and thus should also exhibit a negative and temperature-independent Seebeck coefficient.

Retrospectively, we can understand that the differences in $S/T$ between cleaner cuprates like Nd-LSCO \cite{Gourgout2022Seebeck} and LSCO \cite{jin2021}, and dirtier cuprates like Bi2201~\cite{Kondo2005Contribution} comes only from the differences in impurity concentrations. The residual resistivity in Bi2201 is typically 5 to 20 larger~\cite{kondo_resistivity2006,Ayres2021Incoherent} than in LSCO and Nd-LSCO and the Seebeck coefficient in overdoped Bi2201 is therefore similar to the ones measured in the nickelates (Fig.~\ref{fig:seebeck_bi2201} and Appendix~\ref{app:seebeck_cuprates}).

%#################### Figure 5 ####################%
%##################################################%

Fortuitously, the larger level of elastic disorder in the nickelates makes the Seebeck coefficient entirely insensitive to the scattering rate; similar to the high-field limit of the Hall coefficient, the high-elastic-scattering limit of $S/T$ reflects only the underlying electronic band structure.

The idea that the Seebeck coefficient is solely determined by the electronic band structure in the disordered limit is further supported by two additional examples from the literature: the infinite-layer superconducting nickelate Nd$_{\rm 0.8}$Sr$_{\rm 0.2}$NiO$_2$~\cite{quirk2023} and the delafossite PdCoO$_2$~\cite{Yordanov2019Large}. Quirk \textit{et al.}~\cite{quirk2023} measured the Seebeck effect in Nd$_{\rm 0.8}$Sr$_{\rm 0.2}$NiO$_2$. This film had a residual resistivity of $\rho_{\rm 0} \sim 800$~$\mu\Omega$.cm---similar to our 3-layer nickelate sample and also in the disordered limit. We used Boltzmann transport and a tight-binding model based on the ARPES experiments of Sun \textit{et al.}~\cite{sun_2024} to calculate the Seebeck coefficient. We find excellent agreement between the Seebeck data of Quirk \textit{et al.}~\cite{quirk2023} and calculations in the disordered limit (see Fig.~\ref{fig:seebeck_inf_layer} and Appendix~\ref{app:seebeck_inf_nickelates}). 

Yordanov \textit{et al.}~\cite{Yordanov2019Large} measured the Seebeck effect in thin films of the delafossite PdCoO$_2$. These films were also in the disordered limit, with a residual resistivity 1000 larger than in single-crystal samples. Yordanov \textit{et al.}~\cite{Yordanov2019Large} followed a procedure similar to the one we took for the 5-layer and 3-layer nickelates: they used the band structure from DFT and a standard Boltzmann transport package (that uses elastic scattering by default) to evaluate the Seebeck coefficient. They find perfect agreement between the measurements and the calculations. While Yordanov \textit{et al.}~\cite{Yordanov2019Large} did not connect the success of their calculations to the disordered limit and the constant scattering rate hypothesis, this is another example of the broader validity of our conclusions.

\textbf{Quasiparticles}. $T$-linear and $T+T^2$ resistivity was recently reported down to the lowest temperature in infinite layer nickelates~\cite{lee_linear_character_2023} and Nd$_3$Ni$_2$O$_7$~\cite{sun_2023} under pressure. This suggests a strange-metal phase is present in the nickleates. A $T+T^2$ fit of its resistivity (Fig.~\ref{fig:resistivity_n5})---as standardized in cuprates \cite{Cooper2009Anomalous}---suggests that the $n=5$ nickelate is in proximity to this strange metal regime. While some theories propose that the strange metal regime is a phase without quasiparticles~\cite{hartnoll2021planckian}, the equally good determination of the Seebeck coefficient for both the $n=5$ or $n=3$ nickelates based on a semi-classical Boltzmann transport suggests that this is not the case here. This is in line with the success of several recent studies in cuprates that have demonstrated the validity of the semi-classical approach to describe transport in strange metals~\cite{Grissonnanche2021LinearIn,Fang2022Fermi,Gourgout2022Seebeck,Ataei2022Electrons} and Fermi liquids~\cite{jin2021}. In addition, our study suggests that the band structure of the nickelates as calculated by DFT is a reliable description of the electronic structure of these materials, despite the absence of ARPES measurements to date.

%I think the below is repeteating the same things again

%It also demonstrates the applicability of Boltzmann theory to reliably predict the transport coefficients of quantum materials, even for correlated electron systems.

% >>>>>>>>>>>>>>>>>>>>>>>>>>>>>>>>>>>>>>>>>>>>>>>>>>>>>>>>>>>>>>>>>>>>>>>>>>>>>>>>>>>>>>>>>>>>>>>>>
\section{SUMMARY}

We report the first Seebeck effect study of superconducting nickelates. We used the Seebeck coefficient in the disordered limit to probe the electronic band structure of both a superconducting 5-layer nickelate, as well as metallic 3-layer nickelate. We find the measured Seebeck coefficient is well-described by the band dispersion calculated with DFT, combined with semi-classical transport calculations. The calculated $S/T$ reproduces the amplitude, sign, and temperature dependence of the measured Seebeck coefficient, a rare achievement in predicting transport coefficients in quantum materials, and demonstrates that we have been able to probe the nature of the electronic states in superconducting nickelates---the first report of its kind.

Because of the similar electronic band structures between the nickelates and cuprates, we compare the Seebeck effect in the nickelates with Nd-LSCO -- a cleaner cuprate -- and Bi2201 -- a more disordered cuprate. We find that the Seebeck coefficient for Bi2201 is in perfect agreement with experimental and theoritical data for nickelates. In the case of Nd-LSCO, however, we find a qualitative disagreement despite similarities in the electronic structure of the families. We show that the higher level of disorder present in nickelate thin films and in Bi2201, compared to Nd-LSCO, explains this difference.

As a corollary to our main result, our study highlights that the disordered limit of the Seebeck effect is a powerful and scattering-rate-independent probe of the electronic structure. This is opposite of other transport coefficients like the Hall effect, whose interpretation is opaque in the high scattering rate limit, especially for materials with anisotropic Fermi surfaces like the nickelates. This opens a new avenue of applications for the Seebeck effect in quantum materials by intentionally disordering otherwise-clean materials---for example with electron irradiation---to access intrinsic information about their electronic structure.

% Our results suggest that both semi-classical transport theory and DFT are reliable means for predicting and understanding the transport properties of the nickelates, even in proximity to a strange-metal state. These results also demonstrate the advantage of thermoelectric measurements combined with theoretical tools to probe the electronic dispersion---the same way these tools have met success for describing superconducting cuprates~\cite{Grissonnanche2021LinearIn,Fang2022Fermi,Gourgout2022Seebeck,Ataei2022Electrons}. These recent achievements of uniting theory with experiments to extract fundamental information about the electrons from transport measurements will open a new path to describing the normal state from which unconventional superconductivity emerges in nickelates.

% % %-----------------------------------------%
% % %############# Acknowledgments ###########%
% % %-----------------------------------------%

\section*{Acknowledgments}
% Brad Ramshaw
The work of B.J.R. and G.G. was supported as part of the Institute for Quantum Matter, an Energy Frontier Research Center funded by the U.S. Department of Energy, Office of Science, Basic Energy Sciences under Award No. DE-SC0019331.
 G.A.P. and D.F.S. are primarily supported by U.S. Department of Energy (DOE), Office of Basic Energy Sciences, Division of Materials Sciences and Engineering, under Award No. DE-SC0021925; and by NSF Graduate Research Fellowship Grant No. DGE-1745303.  G.A.P. acknowledges additional support from the Paul \& Daisy Soros Fellowship for New Americans.  Q.S. was supported by the Science and Technology Center for Integrated Quantum Materials, NSF Grant No. DMR-1231319.  J.A.M. acknowledges support from the Packard Foundation and the Gordon and Betty Moore Foundation's EPiQS Initiative, Grant No. GBMF6760.  Materials growth was supported by PARADIM under National Science Foundation (NSF) Cooperative Agreement No. DMR-2039380. We acknowledge the Cornell LASSP Professional Machine Shop for their contributions to designing and fabricating equipment used in this study. H.L and A.S.B acknowledge the support from NSF Grant No. DMR 2045826, the ASU Research Computing Center and the Extreme Science and Engineering Discovery Environment (XSEDE) through research allocation TG-PHY220006, which is supported by NSF grant number ACI-1548562 for HPC resources.

%-----------------------------------------%
%############### Appendix ################%
%-----------------------------------------%

\appendix

\section{DFT Computational Details}
\label{app:dft_details}
Density-functional theory calculations for the $n=5$ and $n=3$ nickelates were performed using the projector augmented plane-wave method as implemented in the VASP code \cite{Kresse1996}. For the exchange-correlation functional, we have used the Perdew-Burke-Ernzerhof (PBE) version of the generalized gradient approximation \cite{Perdew1996generalized}. The reduced Ruddlesden-Popper nickelates crystallize in a tetragonal structure where we have fixed the in-plane lattice constants to match those of the NdGaO$_3$ substrate. The out-of-plane lattice constants were optimized and agree with the experimental values, namely $c = 25.4$ \AA{} and $c = 38.8$ \AA{} for the $n=5$ and $n=3$ materials, respectively \cite{pan_superconductivity_2022}. The size of our plane-wave basis is determined by an energy cutoff of $E_{\mathrm{cut}} = 500$ eV and integration in the Brillouin zone is performed on a $12 \times 12 \times 12$ $k$-mesh for both materials.

\begin{figure*}
  \centering
  \includegraphics[width=\textwidth]{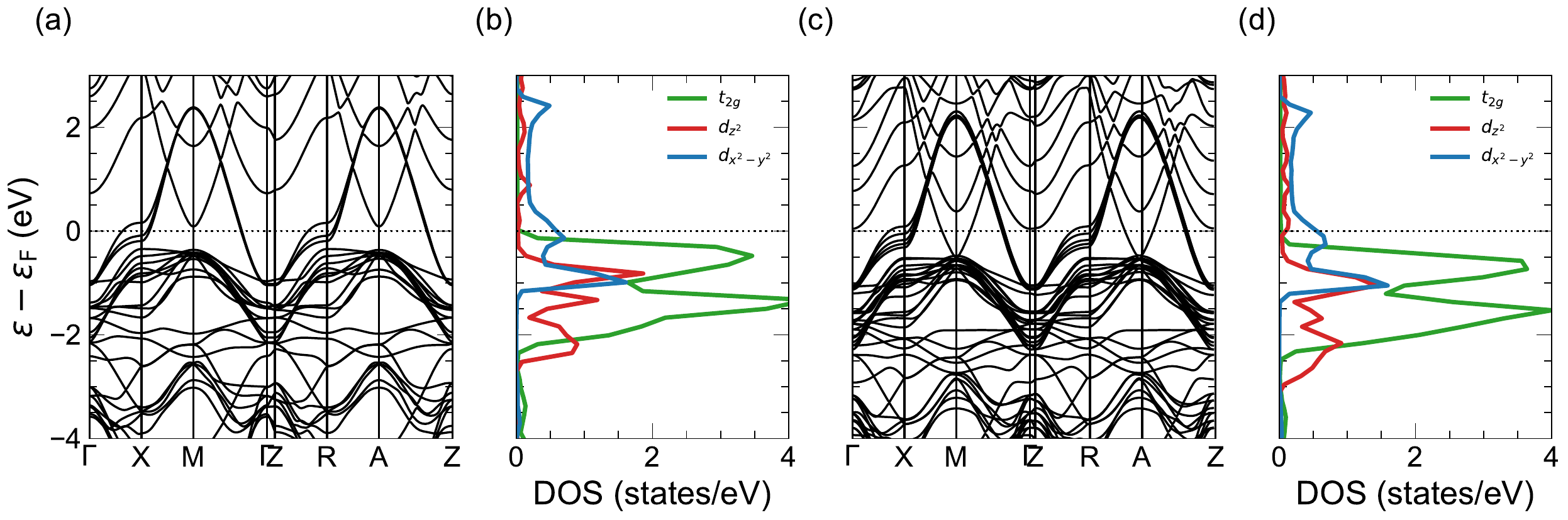}
  \caption{(a) Band structure and (b) orbital-resolved density of states for the Ni($3d$) shell within DFT for the tri-layer nickelate ($n=3$). (c,d) same as (a,b) for the quintuple-layer nickelate ($n=5$), respectively.}
  \label{fig:dft_data}
\end{figure*}

Figure \ref{fig:dft_data} provides a brief summary of the paramagnetic electronic structure of the $n=5$ and $n=3$ nickelates. The band structures reveal a $d_{x^{2}-y^{2}}$ band per NiO$_{2}$ layer crossing the Fermi energy ($E_{\rm F}$), akin to the multi-layer cuprates. Interestingly, for the 5-layer nickelate, there are additional electron pockets at the Brillouin zone corners (M and A) coming from the rare-earth bands. For the 3-layer material, these ``spectator'' bands sit above $E_{\rm F}$. Indeed, the orbital-resolved density of states (DOS) reveals the dominant states are of Ni-$d_{x^{2}-y^{2}}$ character around the $E_{\rm F}$. The Ni-$d_{z^{2}}$ and Ni-$t_{2g}$ ($t_{2g} \equiv \{ d_{xy}, d_{xz}, d_{yz} \}$) states are positioned well below $E_{\rm F}$ and do not play a significant role in the low-energy physics of these materials. For a complete description of the electronic structure of the reduced Ruddlesden-Popper nickelates, see Refs. \onlinecite{Labollita2021electronic, Labollita2022manybody}.

%%%%%%%%%%%%%%%%%%
%%%%%%%%%%%%%%%%%%

%%/////////////////////////////////////////////////////////////////////////////////////
\section{Boltzmann calculations}
\label{app:boltzmann_calc}

\begin{table}
  \begin{center}
    \begin{tabular*}{0.5\textwidth}{c@{\extracolsep{\fill}}cccc}
    \hline\hline
    Band & $\mu/t$ & $t$ (meV) & $t'/t$ & $t''/t$\\
    \hline
    Ni 1 & -1.101 & 396.6 & -0.1833 & 0.1042\\
    Ni 2 & -1.216 & 400.5 & -0.1458 & 0.0855\\
    Ni 3 & -0.765 & 420.9 & -0.2597 & 0.1075\\
    Ni 4 & -0.839 & 425.1 & -0.2483 & 0.0947\\
    Ni 5 & -0.906 & 417.1 & -0.2297 & 0.0795\\
    \hline
    Nd   & 3.157  & -650.0 & 0       & 0     \\
    \hline\hline
    \end{tabular*}
  \end{center}
 \caption{Tight-binding parameters from the bands of $n=5$ nickelate obtained from a fit to the band dispersion calculated by DFT~\cite{Pan2022Synthesis}.}
    \label{tab:tight-binding_5}
\end{table}

\begin{table}
  \begin{center}
    \begin{tabular*}{0.5\textwidth}{c@{\extracolsep{\fill}}cccc}
    \hline\hline
    Band & $\mu/t$ & $t$ (meV) & $t'/t$ & $t''/t$\\
    \hline
    Ni 1 & -1.384 & 410.4 & -0.1532 & 0.0719\\
    Ni 2 & -1.037 & 426.2 & -0.2505 & 0.1071\\
    Ni 3 & -1.138 & 422.1 & -0.2205 & 0.0988\\
    \hline\hline
    \end{tabular*}
  \end{center}
 \caption{Tight-binding parameters from the bands of $n=3$ nickelate obtained from a fit to the band dispersion calculated by DFT~\cite{Pan2022Synthesis}.}
    \label{tab:tight-binding_3}
\end{table}

The Seebeck coefficient is given by the ratio of the Peltier coefficient $\alpha_{ii}$ to the electrical conductivity $\sigma_{ii}$ (with $i = x, z$), $S_i =  \alpha_{ii}/\sigma_{ii}$, where
\begin{equation}
\sigma_{\rm ii}
=
\int_{-\infty}^{\infty} d\epsilon
\Big(-\frac{\partial f(\epsilon)}{\partial \epsilon}\Big)\sigma_{\rm ii}(\epsilon)
\label{eq:sigma}
\end{equation}
\begin{equation}
\alpha_{\rm ii}
=
\int_{-\infty}^{\infty}d\epsilon
\bigg[\Big(-\frac{\partial f(\epsilon)}{\partial \epsilon}\Big)\frac{\epsilon}{T}\bigg]\frac{\sigma_{\rm ii}(\epsilon)}{-e}
\label{eq:alpha}
\end{equation}
with $e$ the electron charge, $f(\epsilon)$ the Fermi-Dirac distribution and
\begin{equation}
\sigma_{\rm ii}(\epsilon)
=
2e^2\iiint_{\text{BZ}}\frac{d^3k}{(2\pi)^3}
v_{\rm i}(\vecck)^2
\tau(\vecck, \epsilon)\delta\big(\epsilon-E(\vecck)\big),
\label{eq:sigma_epsilon}
\end{equation}
where $v_{\rm i}(\vecck)$ is the component of the quasiparticle velocity in the $i$-direction, $\tau(\vecck, \epsilon)$ is the quasiparticle lifetime depending on both momentum $\vecck$ and energy $\epsilon$, and $E(\vecck)$ is given by a tight-binding model.

In order to calculate Seebeck coefficient of the $n=5$ and $n=3$ nickelates, we fitted a tight-binding model $E(\vec k)$ to the band dispersion calculated by DFT with

\begin{equation}
  \begin{aligned}
E(\vec k)=-\mu&-2t[\cos(k_xa)+\cos(k_ya)]
\\&-4t'\cos(k_xa)\cos(k_ya)
\\&-2t''[\cos(2k_xa)+\cos(2k_ya)]
  \end{aligned}
    \label{eq:tight-binding}
\end{equation}

with $a=3.91~\mathring{A}$ ($3.86~\mathring{A}$) and $c=38.8~\mathring{A}$ ($25.4~\mathring{A}$) the lattice constants for the $n=5$ ($n=3$) nickelate. The hopping parameters are found in the tables \ref{tab:tight-binding_5} and \ref{tab:tight-binding_3}.

Two assumptions go into the Boltzmann calculations that have quantitative effects on the calculated value of $S/T$. First, we assume that the scattering rate is the same on all bands. Because the Fermi velocity is of a similar magnitude on all bands, and because the strong elastic scattering is likely dominated by impurities that fix a real-space mean free path, it is reasonable to assume that the elastic mean free path is similar on all bands. This assumption introduces some uncertainty into the absolute value of $S/T$ but does not change it qualitatively as long as the scattering is not radically different (e.g. smaller by a factor of~10 or more) on one of the bands.

Second, we assume that the bandwidth calculated by DFT is the correct one. In real materials, electron-electron interactions tend to lower the overall bandwidth, which in turn reduces our tight-binding bandwidth $t$ and thus increases the calculated $\lvert S/T\rvert$. While a proper measurement of the bandwidth is not available for these films, it is known that DFT has overestimated the bandwidth in lanthanum-based cuprates by about a factor of~2 ~\cite{Markiewicz2005OneBand,Grissonnanche2021LinearIn}. We incorporate a factor of~2 uncertainty in the bandwidth into our calculated $S/T$ in Fig.~\ref{fig:dft_seebeck}d~and~e.
%%/////////////////////////////////////////////////////////////////////////////////////
\section{Sample comparison of $n=5$ nickelates}
\label{app:sample_comparison}

\begin{figure}
  \includegraphics[width=0.42\textwidth]{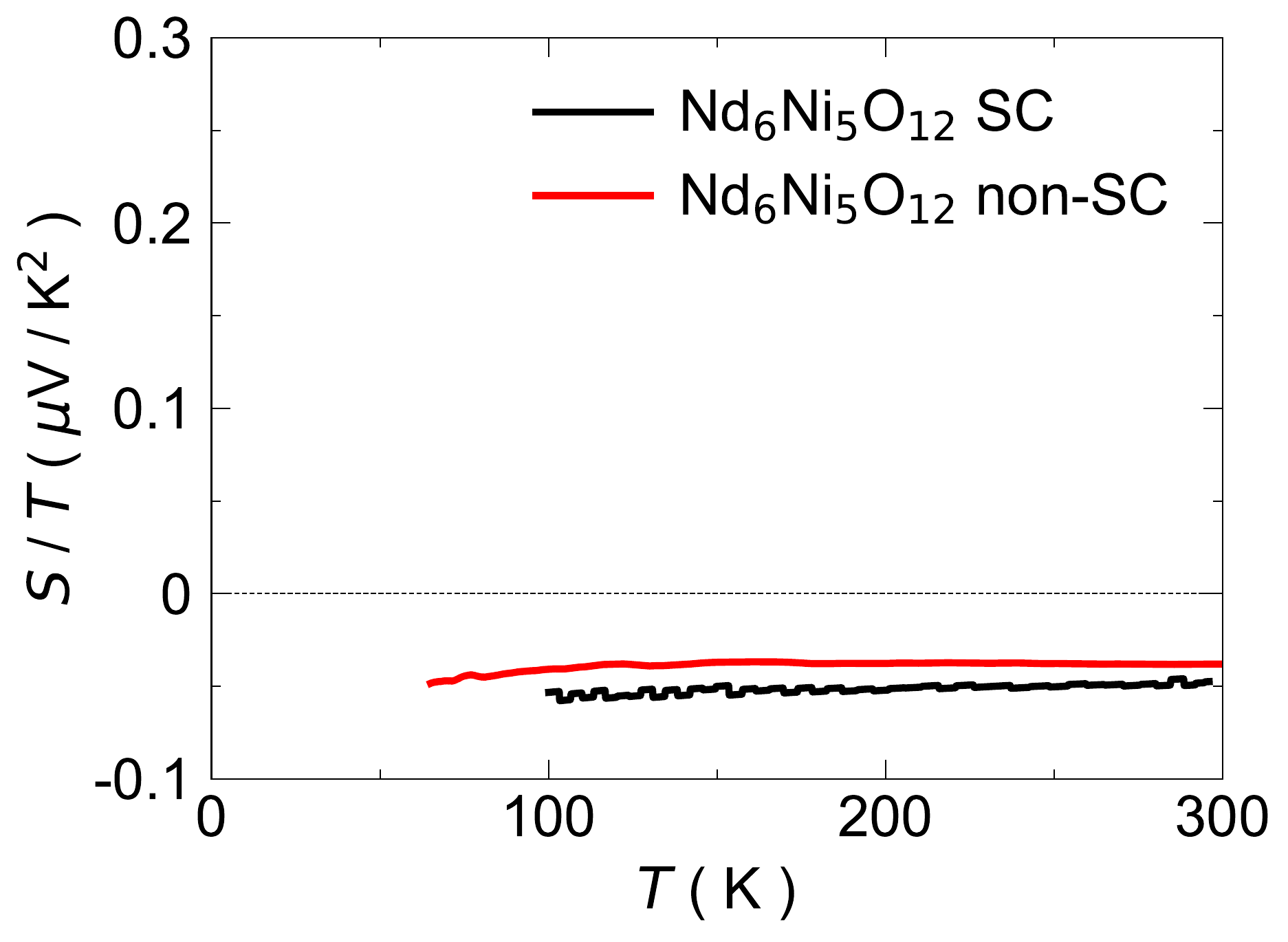}
  \caption{$S/T$ as a function of temperature of two different $n=5$ nickelate samples. The first sample is a superconducting nickelate thin film with a substrate thickness of $500$~microns. The non-superconducting nickelate has a reduced substrate thickness down to $150$~microns.}
  \label{fig:sample_comparison}
  \end{figure}

Here we compare the Seebeck coefficient of two samples of $n=5$ nickelate (Fig.~\ref{fig:sample_comparison}). The superconducting sample was only measured down to 100~kelvin due to having a thicker substrate, which made it impossible within the temperature resolution to generate a sizable thermal gradient below that 100~kelvin to measure the Seebeck effect. Indeed, the thermal conductivity of the substrate, NdGaO$_3$, increases dramatically at low temperature, short-circuiting any attempt to generate a thermal gradient with a reasonable amount of heat. The second sample was grown in similar conditions, but did not exhibit superconductivity due to the sensitivity of the superconducting state to few percent changes in the cation stoichiometry. We were able to reduce the thickness of that sample substrate down to $150$~microns to be able to measure the Seebeck effect down to lower temperature $\approx 60$~K on this sample. Fortunately, the Seebeck coefficients between the two samples are very similar and agree to within 15\%.  This difference may be accounted for by the varying levels of cation disorder introduced during the MBE-synthesis of the two samples, as well as by the randomness inherent to the chemical reduction process used in the synthesis of all square-planar nickelates. Nevertheless, the overall reproducibility confirms that the normal state is similar between these samples.

%%%%%%%%%%%%%%%%%%
  %%%%%%%%%%%%%%%%%%

%%/////////////////////////////////////////////////////////////////////////////////////
\section{Rare-earth band}
\label{app:rare-earth}

One significant difference between the two nickelates is the presence of a neodymium band crossing the Fermi level for the superconducting, 5-layer nickelate. The role of this band is unknown---whether it contributes significantly to the conductivity, or even to the superconducting pairing~\cite{Louie2022}. To include the neodymium band in the Boltzmann transport calculations changes from $S/T = -37.2$~nV~/~K$^2$ with it to $S/T = -33.4$~nV~/~K$^2$ without. This 10\% difference is likely to remain undetected within the experimental error bars. Therefore, it is difficult to conclude whether the rare-earth band participates in the measured Seebeck coefficient as calculations indicate its contribution remains marginal. This could suggest that the neodymium band does not play a dominant role in the metallic state of the 5-layer superconducting nickelate, which in turn suggests that it may not play a role in the superconductivity.

% %%//////////////////////////////////////////////////////////////////////////////
% \section{Elastic scattering dependence}
% \label{app:elastic_dependence}

% In the calculations of the Seebeck coefficient of Nd-LSCO $p=0.24$, the total scattering rate by $1/\tau = 1/\tau_{\rm 0} + 1/\tau(\epsilon)$. The Seebeck coefficient changes from positive at small values of $1/\tau_{\rm 0}$ to negative and independent of the elastic scattering rate for larger values $1/\tau_{\rm 0}$ as shown in Figure~\ref{fig:impurity_appendix}.

% %%%%%%%%%%%%%%%%%%
% \begin{figure}
%   \includegraphics[width=0.42\textwidth]{Fig_S_many_tau0.pdf}
%   \caption{Calculated Seebeck coefficient, plotted as $S/T$ of Nd-LSCO $p=0.24$ as a function of elastic scattering rate at $T=4$~K. The total scattering rate is given by the $1/\tau_{\rm 0} + 1/\tau(\epsilon)$.}
%   \label{fig:impurity_appendix}
%   \end{figure}
%   %%%%%%%%%%%%%%%%%%

%%/////////////////////////////////////////////////////////////////////////////////////
\section{Resistivity of the nickelates}
\label{app:resistivity}

%%%%%%%%%%%%%%%%%%
\begin{figure}[h]
\includegraphics[width=0.35\textwidth]{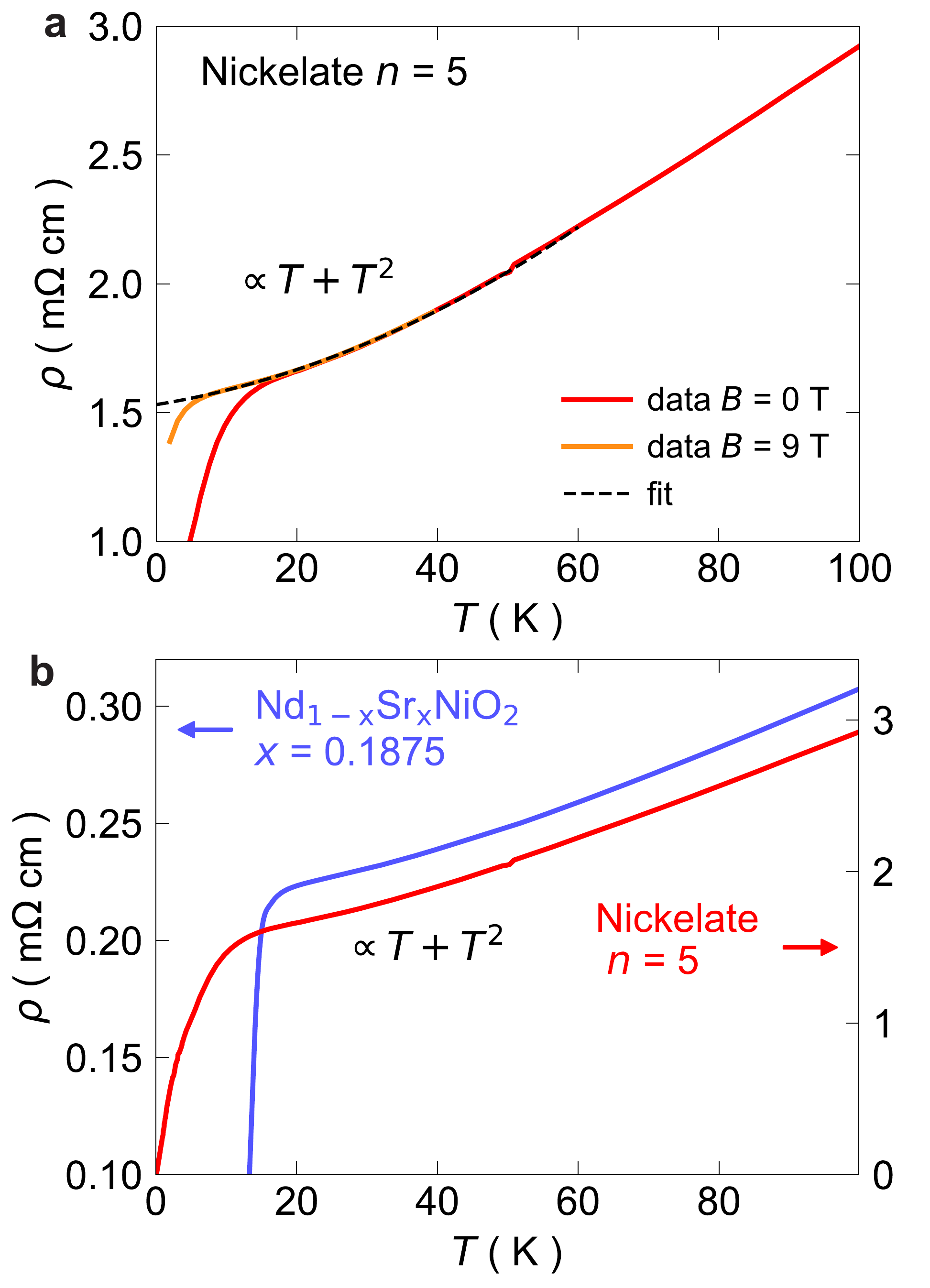}
\caption{Resistivity $\rho$ as a function of temperature for (\textbf{a}) $n=5$ nickelate in $B$ = 0 and 9~T, and (\textbf{b}) $n=5$ nickelate and infinite layer nickelate Nd$_{\rm 1-x}$Sr$_{\rm x}$NiO$_{\rm 2}$ $x=0.1875$ in $B=0$~\cite{lee_linear_character_2023}. Both materials are in the strange metal regime, as demonstrated by their $T+T^2$ resistivity.}
\label{fig:resistivity_n5}
\end{figure}
%%%%%%%%%%%%%%%%%%

% %%%%%%%%%%%%%%%%%%
% \begin{figure}[h]
% \includegraphics[width=0.35\textwidth]{Fig_rho_n_3_fit.pdf}
% \caption{Resistivity $\rho$ as a function of temperature for $n=3$ nickelate in $B=0$. The faded upturn part of the resistivity for $n=3$ betrays signatures of weak localization due to disorder at low temperature, as observed so far in all overdoped nickelates~\cite{lee_linear_character_2023}, and was not part of the fit.}
% \label{fig:resistivity_n3}
% \end{figure}
% %%%%%%%%%%%%%%%%%%

While $T$-linear resistivity at high temperature is found in conventional metals like copper, because electrons scatter quasi-elastically off of phonons, this mechanism fades away at low temperature and this behavior does not persist down to $T=0$. The term ``strange metal'' within the community of strongly correlated electron systems describes a metal whose resistivity remains linear in temperature down to $T = 0$~\cite{Bruin2013Similarity,Legros2019Universal}, an inexplicable behavior to date. Here we focus on the $T$-linear component of the resistivity in the $T\rightarrow 0$ limit because it is unambiguously strange.

There are two caveats to demonstrate that the $n = 5$ nickelate is a strange metal. First, superconductivity masks the $T = 0$ limit of the resistivity. For this reason, we fit the resistivity at $B=9$~T down to the lowest temperature above $T_c$. Reaching a higher field is difficult, as the sample tends to move in a magnetic field because of the torque in the substrate NdGaO$_3$. Second, the resistivity is not purely linear in temperature nor is it purely quadratic. However, claiming that the $n = 5$ nickelate is in the strange metal regime of the nickelates’ phase diagram comes from its $T + T^2$ resistivity fitted over a decade from 6~K to 60~K. The definition of the “strange metal regime” used in cuprates~\cite{Cooper2009Anomalous}, iron-based \cite{fang2009} and organic superconductors~\cite{doiron-leyraud2009}, consists of the observation of a $T + T^2$ resistivity with a significant $T$-linear component that persists down to $T = 0$. Perfectly $T$-linear resistivity is only observed at a particular doping in these materials---away from this doping, the resistivity gains a $T^2$ component.

The strange metal regime has now also been reported in doped, infinite-layer nickelates with $T$-linear, $T + T^2 $ and $T^2$ resistivities~\cite{lee_linear_character_2023} (Fig.~\ref{fig:resistivity_n5}b) and Nd$_3$Ni$_2$O$_7$ \cite{sun_2023}. Therefore, the strange metal regime in the nickelates follows the same doping dependence observed in the cuprates, iron-based and organic superconductors.

The fit of the resistivity of the $n=5$ nickelate sample includes a significant $T$-linear component and can be described by $\rho(T) = \rho_0 + a_1 T + a_2 T^2$ as shown in Figure~\ref{fig:resistivity_n5}a, with values $\rho_0=1450$~$\mu\Omega$cm, $a_1=8.1$~$\mu\Omega$cm~/~K, and $a_2=0.0695$~$\mu\Omega$cm~/~K$^2$. And by comparison, the temperature dependence of the resistivity is very similar to the one of Nd$_{\rm 1-x}$Sr$_{\rm x}$NiO$_{\rm 2}$ $x = 0.1875$~\cite{lee_linear_character_2023}, a doping with $T+T^2$ behavior (Fig.~\ref{fig:resistivity_n5}).

% In contrast, the $n=3$ nickelate’s resistivity can be fit purely to $T^2$ and shows no sign of a $T$-linear component, as expected for a Fermi liquid and for nickelates (and cuprates) the more overdoped region of the phase diagram. In this case, $\rho(T) = \rho_0 + a_1 T + a_2 T^2$ as shown in Figure~\ref{fig:resistivity_n3}b, with values $\rho_0=920$~$\mu\Omega$cm, $a_1=0.1$~$\mu\Omega$cm~/~K, and $a_2=0.0739$~$\mu\Omega$cm~/~K$^2$. It is, therefore, not a strange metal and should be compared to very overdoped Nd$_{\rm 1-x}$Sr$_{\rm x}$NiO$_{\rm 2}$ $x = 0.275$~\cite{lee_linear_character_2023}.}

%%/////////////////////////////////////////////////////////////////////////////////////
\section{Seebeck coefficient in infinite-layer nickelates}
\label{app:seebeck_inf_nickelates}
The Seebeck coefficient was recently measured in the infinite-layer nickelate Nd$_{\rm 1-x}$Sr$_{\rm x}$NiO$_2$~\cite{quirk2023} at doping $x=0.2$. The measured film has a large residual resistivity of $\rho_0 \sim 800$ m$\Omega$.cm, which, as we will show below, places it in the same disordered limit as the $n=5$- and $n=3$-layer nickelates.

%%%%%%%%%%%%%%%%%%
\begin{figure}[h]
\includegraphics[width=0.5\textwidth]{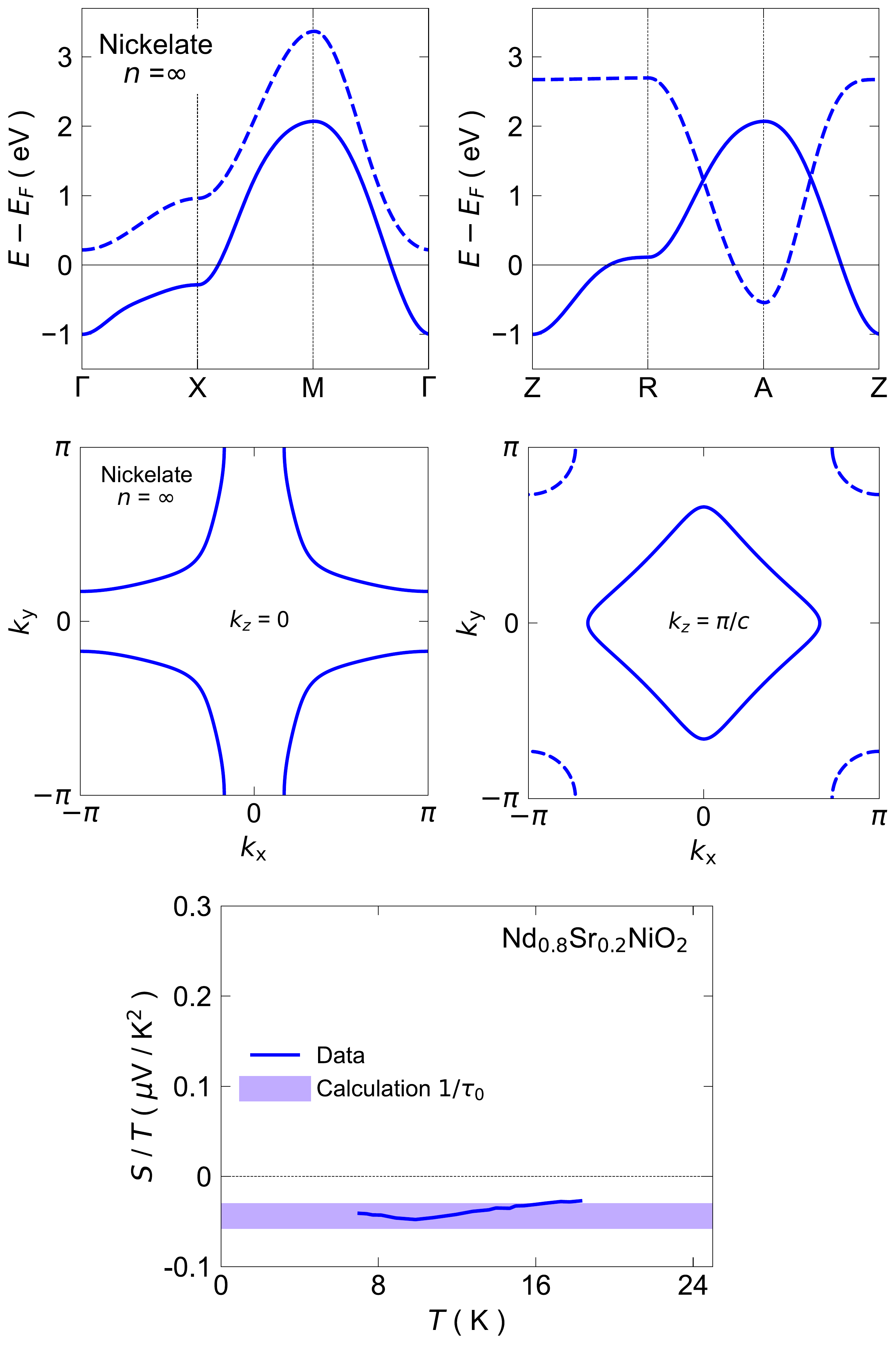}
\caption{Electronic band structure (top), Fermi surface (middle) and calculated Seebeck coefficient plotted as $S/T$ vs $T$ (bottom) for the infinite-layer nickelate RE$_{\rm 0.8}$Sr$_{\rm 0.2}$NiO$_2$ at doping $x=0.2$, with RE = La or Nd. The electronic band structure and the Fermi surface was measured by ARPES~\cite{sun_2024} on La$_{\rm 0.8}$Sr$_{\rm 0.2}$NiO$_2$. It includes two sheets: a large Ni $d_{\rm x^2-y^2}$ sheet that evolves from hole-like to electron-like along $k_z$ (full line), and a three-dimensional electron pocket centered at Brillouin zone corner (dashed line). The bottom panel compares the measured Seebeck coefficient on Nd$_{\rm 0.8}$Sr$_{\rm 0.2}$NiO$_2$~\cite{quirk2023} to the one calculated from the band structure using Boltzmann transport and a constant elastic scattering rate $1/\tau_0$.
}
\label{fig:seebeck_inf_layer}
\end{figure}
%%%%%%%%%%%%%%%%%%

We use the electronic dispersion obtained from recent ARPES measurement on La$_{\rm 1-x}$Sr$_{\rm x}$NiO$_2$ at doping $x~=~0.2$~\cite{sun_2024}---the same doping measured in the Seebeck experiments---with an identical crystal structure but with a different rare-earth that contributes only marginally to the transport (see Appendix~\ref{app:rare-earth}). Here, Sun \textit{et al.}~\cite{sun_2024} was able to show that the ARPES electronic dispersion is in good agreement with the DFT calculations on the same material, corresponding to the hybridization of a nickel d-band with a La-band (Fig.~\ref{fig:seebeck_inf_layer} top).

Combining the ARPES electronic dispersion and Boltzmann transport, we compute $S/T$ and find a value in excellent agreement with the experiment (Fig.~\ref{fig:seebeck_inf_layer} bottom), similar to what we found for the $n=5$-, $n=3$-layer nickelates, and Bi2201. By computing the resistivity, we are able to show that the residual resistivity of  $\rho_0 \sim 800$ m$\Omega$.cm from \citet{quirk2023} corresponds to a scattering rate $1/\tau_0 \sim 3000$~ps$^-1$. Fig.~\ref{fig:impurity}b places it in the same ballpark as $n=5$-layer nickelate, \textit{i.e.} in the disordered limit. This further supports our conclusion that Seebeck measurements in the disordered limit are sensitive only to the shape of the electronic band structure.

%%/////////////////////////////////////////////////////////////////////////////////////
\section{Seebeck coefficient in cuprates}
\label{app:seebeck_cuprates}
The doping range accessible can vary enormously from a cuprate compound to another, and finding a cuprate sample that shares the same electronic phase as the $n=5$ and $n=3$ nickelates and whose Seebeck coefficient has been measured is not a trivial task. This is why we chose the comparison with Nd-LSCO $p=0.24$ and Bi2201 $p=0.23$, which are metallic and free from the pseudogap phase and charge order, similarly to the nickelates.

%%%%%%%%%%%%%%%%%%
\begin{figure}[b!]
\includegraphics[width=0.42\textwidth]{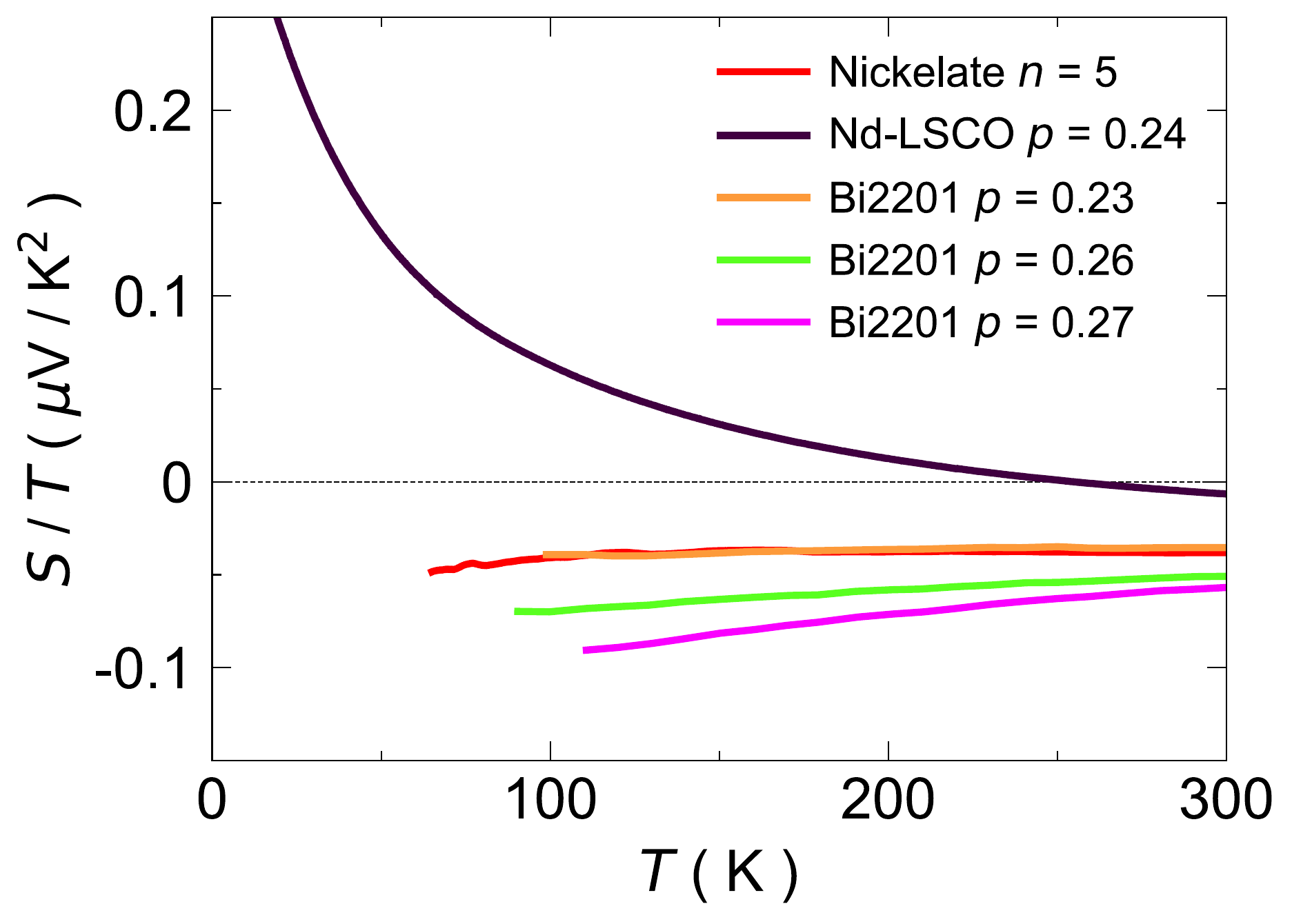}
\caption{In-plane Seebeck coefficient plotted as $S/T$ vs $T$ of Nd-LSCO $p=0.24$ \cite{Gourgout2022Seebeck}, $n=5$ nickelate and Bi2201 at doping $p=0.23$, $0.26$, $0.27$.}
\label{fig:seebeck_bi2201}
\end{figure}
%%%%%%%%%%%%%%%%%%

%Regarding Nd-LSCO, the material cannot be grown much more overdoped than $p = 0.24$ \cite{dragomir2020}, and the Seebeck effect has not been measured beyond this doping. However, 
The Seebeck effect has been measured for $p<0.24$ as well in Nd-LSCO by Gourgout \textit{et al.}~\cite{Gourgout2022Seebeck}, where Nd-LSCO is in the pseudogap phase with a Fermi surface transformation happening at $p^*=0.23$ \cite{Fang2022Fermi}. Below $p^*$, the behavior of $S/T$ remains similar to that of Nd-LSCO $p = 0.24$, meaning positive and strongly temperature dependent, with a larger amplitude.

Badoux \textit{et al.} \cite{badoux2016} measured the Seebeck effect in LSCO between $p = 0.07$ and $p = 0.15$---a very different regime than Nd-LSCO $p=0.24$ and the nickelates. Above the charge density wave onset temperature, $S/T$ is temperature dependent and positive like Nd-LSCO. However, below the charge ordering temperature, $S/T$ becomes negative and remains strongly temperature dependent down to $T = 0$. In contrast, in Jin~\textit{et al.}~\cite{jin2021}, authors report Seebeck data on very overdoped LSCO $p = 0.33$, doping where the resistivity is quadratic in temperature and the Seebeck coefficient is qualitatively similar to Nd-LSCO $p = 0.24$ but with a lower amplitude.

The closest comparison to the Seebeck effect found in the nickelates in this study is in the overdoped cuprate (Bi,Pb)$_{\rm 2}$(Sr,La)$_{\rm 2}$CuO$_{\rm 6+\delta}$ (Bi2201) \cite{Kondo2005Contribution}. At high doping $p > 0.23$, the Seebeck effect $S$ is reported linear in temperature and negative---exactly as we find in the nickelates. We show this striking comparison in Fig.~\ref{fig:seebeck_bi2201}. Bi2201 is significantly more disordered than LSCO and Nd-LSCO, with typically residual resistivities $\rho_{\rm 0} \sim 120$~$\mu\Omega$ cm or more \cite{Ayres2021Incoherent}, which is 5-20 times larger than LSCO and Nd-LSCO, meaning the samples come with much higher elastic scattering from defects.

% %-----------------------------------------%
% %############### Biblio ##################%
% %-----------------------------------------%

% % Create the reference section using BibTeX:
% \bibliography{references}

\begin{thebibliography}{49}%
\makeatletter
\providecommand \@ifxundefined [1]{%
 \@ifx{#1\undefined}
}%
\providecommand \@ifnum [1]{%
 \ifnum #1\expandafter \@firstoftwo
 \else \expandafter \@secondoftwo
 \fi
}%
\providecommand \@ifx [1]{%
 \ifx #1\expandafter \@firstoftwo
 \else \expandafter \@secondoftwo
 \fi
}%
\providecommand \natexlab [1]{#1}%
\providecommand \enquote  [1]{``#1''}%
\providecommand \bibnamefont  [1]{#1}%
\providecommand \bibfnamefont [1]{#1}%
\providecommand \citenamefont [1]{#1}%
\providecommand \href@noop [0]{\@secondoftwo}%
\providecommand \href [0]{\begingroup \@sanitize@url \@href}%
\providecommand \@href[1]{\@@startlink{#1}\@@href}%
\providecommand \@@href[1]{\endgroup#1\@@endlink}%
\providecommand \@sanitize@url [0]{\catcode `\\12\catcode `\$12\catcode
  `\&12\catcode `\#12\catcode `\^12\catcode `\_12\catcode `\%12\relax}%
\providecommand \@@startlink[1]{}%
\providecommand \@@endlink[0]{}%
\providecommand \url  [0]{\begingroup\@sanitize@url \@url }%
\providecommand \@url [1]{\endgroup\@href {#1}{\urlprefix }}%
\providecommand \urlprefix  [0]{URL }%
\providecommand \Eprint [0]{\href }%
\providecommand \doibase [0]{https://doi.org/}%
\providecommand \selectlanguage [0]{\@gobble}%
\providecommand \bibinfo  [0]{\@secondoftwo}%
\providecommand \bibfield  [0]{\@secondoftwo}%
\providecommand \translation [1]{[#1]}%
\providecommand \BibitemOpen [0]{}%
\providecommand \bibitemStop [0]{}%
\providecommand \bibitemNoStop [0]{.\EOS\space}%
\providecommand \EOS [0]{\spacefactor3000\relax}%
\providecommand \BibitemShut  [1]{\csname bibitem#1\endcsname}%
\let\auto@bib@innerbib\@empty
%</preamble>
\bibitem [{\citenamefont {Keimer}\ \emph {et~al.}(2015)\citenamefont {Keimer},
  \citenamefont {Kivelson}, \citenamefont {Norman}, \citenamefont {Uchida},\
  and\ \citenamefont {Zaanen}}]{Keimer2015From}%
  \BibitemOpen
  \bibfield  {author} {\bibinfo {author} {\bibfnamefont {B.}~\bibnamefont
  {Keimer}}, \bibinfo {author} {\bibfnamefont {S.~A.}\ \bibnamefont
  {Kivelson}}, \bibinfo {author} {\bibfnamefont {M.~R.}\ \bibnamefont
  {Norman}}, \bibinfo {author} {\bibfnamefont {S.}~\bibnamefont {Uchida}},\
  and\ \bibinfo {author} {\bibfnamefont {J.}~\bibnamefont {Zaanen}},\
  }\bibfield  {title} {\bibinfo {title} {From quantum matter to
  high-temperature superconductivity in copper oxides},\ }\href
  {https://doi.org/10.1038/nature14165} {\bibfield  {journal} {\bibinfo
  {journal} {Nature}\ }\textbf {\bibinfo {volume} {518}},\ \bibinfo {pages}
  {179} (\bibinfo {year} {2015})}\BibitemShut {NoStop}%
\bibitem [{\citenamefont {Maeno}\ \emph {et~al.}(1994)\citenamefont {Maeno},
  \citenamefont {Hashimoto}, \citenamefont {Yoshida}, \citenamefont
  {Nishizaki}, \citenamefont {Fujita}, \citenamefont {Bednorz},\ and\
  \citenamefont {Lichtenberg}}]{maeno_superconductivity_1994}%
  \BibitemOpen
  \bibfield  {author} {\bibinfo {author} {\bibfnamefont {Y.}~\bibnamefont
  {Maeno}}, \bibinfo {author} {\bibfnamefont {H.}~\bibnamefont {Hashimoto}},
  \bibinfo {author} {\bibfnamefont {K.}~\bibnamefont {Yoshida}}, \bibinfo
  {author} {\bibfnamefont {S.}~\bibnamefont {Nishizaki}}, \bibinfo {author}
  {\bibfnamefont {T.}~\bibnamefont {Fujita}}, \bibinfo {author} {\bibfnamefont
  {J.~G.}\ \bibnamefont {Bednorz}},\ and\ \bibinfo {author} {\bibfnamefont
  {F.}~\bibnamefont {Lichtenberg}},\ }\bibfield  {title} {\bibinfo {title}
  {Superconductivity in a layered perovskite without copper},\ }\bibfield
  {journal} {\bibinfo  {journal} {Nature}\ }\textbf {\bibinfo {volume} {372}},\
  \href {https://doi.org/10.1038/372532a0} {10.1038/372532a0} (\bibinfo {year}
  {1994})\BibitemShut {NoStop}%
\bibitem [{\citenamefont {Li}\ \emph {et~al.}(2019)\citenamefont {Li},
  \citenamefont {Lee}, \citenamefont {Wang}, \citenamefont {Osada},
  \citenamefont {Crossley}, \citenamefont {Lee}, \citenamefont {Cui},
  \citenamefont {Hikita},\ and\ \citenamefont
  {Hwang}}]{Li2019Superconductivity}%
  \BibitemOpen
  \bibfield  {author} {\bibinfo {author} {\bibfnamefont {D.}~\bibnamefont
  {Li}}, \bibinfo {author} {\bibfnamefont {K.}~\bibnamefont {Lee}}, \bibinfo
  {author} {\bibfnamefont {B.~Y.}\ \bibnamefont {Wang}}, \bibinfo {author}
  {\bibfnamefont {M.}~\bibnamefont {Osada}}, \bibinfo {author} {\bibfnamefont
  {S.}~\bibnamefont {Crossley}}, \bibinfo {author} {\bibfnamefont {H.~R.}\
  \bibnamefont {Lee}}, \bibinfo {author} {\bibfnamefont {Y.}~\bibnamefont
  {Cui}}, \bibinfo {author} {\bibfnamefont {Y.}~\bibnamefont {Hikita}},\ and\
  \bibinfo {author} {\bibfnamefont {H.~Y.}\ \bibnamefont {Hwang}},\ }\bibfield
  {title} {\bibinfo {title} {Superconductivity in an infinite-layer
  nickelate},\ }\href {https://doi.org/10.1038/s41586-019-1496-5} {\bibfield
  {journal} {\bibinfo  {journal} {Nature}\ }\textbf {\bibinfo {volume} {572}},\
  \bibinfo {pages} {624} (\bibinfo {year} {2019})}\BibitemShut {NoStop}%
\bibitem [{\citenamefont {Li}\ \emph {et~al.}(2020)\citenamefont {Li},
  \citenamefont {Wang}, \citenamefont {Lee}, \citenamefont {Harvey},
  \citenamefont {Osada}, \citenamefont {Goodge}, \citenamefont {Kourkoutis},\
  and\ \citenamefont {Hwang}}]{Danfeng2020}%
  \BibitemOpen
  \bibfield  {author} {\bibinfo {author} {\bibfnamefont {D.}~\bibnamefont
  {Li}}, \bibinfo {author} {\bibfnamefont {B.~Y.}\ \bibnamefont {Wang}},
  \bibinfo {author} {\bibfnamefont {K.}~\bibnamefont {Lee}}, \bibinfo {author}
  {\bibfnamefont {S.~P.}\ \bibnamefont {Harvey}}, \bibinfo {author}
  {\bibfnamefont {M.}~\bibnamefont {Osada}}, \bibinfo {author} {\bibfnamefont
  {B.~H.}\ \bibnamefont {Goodge}}, \bibinfo {author} {\bibfnamefont {L.~F.}\
  \bibnamefont {Kourkoutis}},\ and\ \bibinfo {author} {\bibfnamefont {H.~Y.}\
  \bibnamefont {Hwang}},\ }\bibfield  {title} {\bibinfo {title}
  {Superconducting dome in
  ${\mathrm{nd}}_{1\ensuremath{-}x}{\mathrm{sr}}_{x}{\mathrm{nio}}_{2}$
  infinite layer films},\ }\href
  {https://doi.org/10.1103/PhysRevLett.125.027001} {\bibfield  {journal}
  {\bibinfo  {journal} {Phys. Rev. Lett.}\ }\textbf {\bibinfo {volume} {125}},\
  \bibinfo {pages} {027001} (\bibinfo {year} {2020})}\BibitemShut {NoStop}%
\bibitem [{\citenamefont {Zeng}\ \emph {et~al.}(2020)\citenamefont {Zeng},
  \citenamefont {Tang}, \citenamefont {Yin}, \citenamefont {Li}, \citenamefont
  {Li}, \citenamefont {Huang}, \citenamefont {Hu}, \citenamefont {Liu},
  \citenamefont {Omar}, \citenamefont {Jani}, \citenamefont {Lim},
  \citenamefont {Han}, \citenamefont {Wan}, \citenamefont {Yang}, \citenamefont
  {Pennycook}, \citenamefont {Wee},\ and\ \citenamefont {Ariando}}]{Zeng2020}%
  \BibitemOpen
  \bibfield  {author} {\bibinfo {author} {\bibfnamefont {S.}~\bibnamefont
  {Zeng}}, \bibinfo {author} {\bibfnamefont {C.~S.}\ \bibnamefont {Tang}},
  \bibinfo {author} {\bibfnamefont {X.}~\bibnamefont {Yin}}, \bibinfo {author}
  {\bibfnamefont {C.}~\bibnamefont {Li}}, \bibinfo {author} {\bibfnamefont
  {M.}~\bibnamefont {Li}}, \bibinfo {author} {\bibfnamefont {Z.}~\bibnamefont
  {Huang}}, \bibinfo {author} {\bibfnamefont {J.}~\bibnamefont {Hu}}, \bibinfo
  {author} {\bibfnamefont {W.}~\bibnamefont {Liu}}, \bibinfo {author}
  {\bibfnamefont {G.~J.}\ \bibnamefont {Omar}}, \bibinfo {author}
  {\bibfnamefont {H.}~\bibnamefont {Jani}}, \bibinfo {author} {\bibfnamefont
  {Z.~S.}\ \bibnamefont {Lim}}, \bibinfo {author} {\bibfnamefont
  {K.}~\bibnamefont {Han}}, \bibinfo {author} {\bibfnamefont {D.}~\bibnamefont
  {Wan}}, \bibinfo {author} {\bibfnamefont {P.}~\bibnamefont {Yang}}, \bibinfo
  {author} {\bibfnamefont {S.~J.}\ \bibnamefont {Pennycook}}, \bibinfo {author}
  {\bibfnamefont {A.~T.~S.}\ \bibnamefont {Wee}},\ and\ \bibinfo {author}
  {\bibfnamefont {A.}~\bibnamefont {Ariando}},\ }\bibfield  {title} {\bibinfo
  {title} {Phase diagram and superconducting dome of infinite-layer
  ${\mathrm{nd}}_{1\ensuremath{-}x}{\mathrm{sr}}_{x}{\mathrm{nio}}_{2}$ thin
  films},\ }\href {https://doi.org/10.1103/PhysRevLett.125.147003} {\bibfield
  {journal} {\bibinfo  {journal} {Phys. Rev. Lett.}\ }\textbf {\bibinfo
  {volume} {125}},\ \bibinfo {pages} {147003} (\bibinfo {year}
  {2020})}\BibitemShut {NoStop}%
\bibitem [{\citenamefont {Pan}\ \emph {et~al.}(2022{\natexlab{a}})\citenamefont
  {Pan}, \citenamefont {Ferenc~Segedin}, \citenamefont {LaBollita},
  \citenamefont {Song}, \citenamefont {Nica}, \citenamefont {Goodge},
  \citenamefont {Pierce}, \citenamefont {Doyle}, \citenamefont {Novakov},
  \citenamefont {Córdova~Carrizales}, \citenamefont {N’Diaye}, \citenamefont
  {Shafer}, \citenamefont {Paik}, \citenamefont {Heron}, \citenamefont {Mason},
  \citenamefont {Yacoby}, \citenamefont {Kourkoutis}, \citenamefont {Erten},
  \citenamefont {Brooks}, \citenamefont {Botana},\ and\ \citenamefont
  {Mundy}}]{pan_superconductivity_2022}%
  \BibitemOpen
  \bibfield  {author} {\bibinfo {author} {\bibfnamefont {G.~A.}\ \bibnamefont
  {Pan}}, \bibinfo {author} {\bibfnamefont {D.}~\bibnamefont {Ferenc~Segedin}},
  \bibinfo {author} {\bibfnamefont {H.}~\bibnamefont {LaBollita}}, \bibinfo
  {author} {\bibfnamefont {Q.}~\bibnamefont {Song}}, \bibinfo {author}
  {\bibfnamefont {E.~M.}\ \bibnamefont {Nica}}, \bibinfo {author}
  {\bibfnamefont {B.~H.}\ \bibnamefont {Goodge}}, \bibinfo {author}
  {\bibfnamefont {A.~T.}\ \bibnamefont {Pierce}}, \bibinfo {author}
  {\bibfnamefont {S.}~\bibnamefont {Doyle}}, \bibinfo {author} {\bibfnamefont
  {S.}~\bibnamefont {Novakov}}, \bibinfo {author} {\bibfnamefont
  {D.}~\bibnamefont {Córdova~Carrizales}}, \bibinfo {author} {\bibfnamefont
  {A.~T.}\ \bibnamefont {N’Diaye}}, \bibinfo {author} {\bibfnamefont
  {P.}~\bibnamefont {Shafer}}, \bibinfo {author} {\bibfnamefont
  {H.}~\bibnamefont {Paik}}, \bibinfo {author} {\bibfnamefont {J.~T.}\
  \bibnamefont {Heron}}, \bibinfo {author} {\bibfnamefont {J.~A.}\ \bibnamefont
  {Mason}}, \bibinfo {author} {\bibfnamefont {A.}~\bibnamefont {Yacoby}},
  \bibinfo {author} {\bibfnamefont {L.~F.}\ \bibnamefont {Kourkoutis}},
  \bibinfo {author} {\bibfnamefont {O.}~\bibnamefont {Erten}}, \bibinfo
  {author} {\bibfnamefont {C.~M.}\ \bibnamefont {Brooks}}, \bibinfo {author}
  {\bibfnamefont {A.~S.}\ \bibnamefont {Botana}},\ and\ \bibinfo {author}
  {\bibfnamefont {J.~A.}\ \bibnamefont {Mundy}},\ }\bibfield  {title} {\bibinfo
  {title} {Superconductivity in a quintuple-layer square-planar nickelate},\
  }\href {https://doi.org/10.1038/s41563-021-01142-9} {\bibfield  {journal}
  {\bibinfo  {journal} {Nature Materials}\ }\textbf {\bibinfo {volume} {21}},\
  \bibinfo {pages} {160} (\bibinfo {year} {2022}{\natexlab{a}})}\BibitemShut
  {NoStop}%
\bibitem [{\citenamefont {Poltavets}\ \emph {et~al.}(2007)\citenamefont
  {Poltavets}, \citenamefont {Lokshin}, \citenamefont {Croft}, \citenamefont
  {Mandal}, \citenamefont {Egami},\ and\ \citenamefont
  {Greenblatt}}]{Poltavets2007crystal}%
  \BibitemOpen
  \bibfield  {author} {\bibinfo {author} {\bibfnamefont {V.~V.}\ \bibnamefont
  {Poltavets}}, \bibinfo {author} {\bibfnamefont {K.~A.}\ \bibnamefont
  {Lokshin}}, \bibinfo {author} {\bibfnamefont {M.}~\bibnamefont {Croft}},
  \bibinfo {author} {\bibfnamefont {T.~K.}\ \bibnamefont {Mandal}}, \bibinfo
  {author} {\bibfnamefont {T.}~\bibnamefont {Egami}},\ and\ \bibinfo {author}
  {\bibfnamefont {M.}~\bibnamefont {Greenblatt}},\ }\bibfield  {title}
  {\bibinfo {title} {Crystal structures of ln4ni3o8 (ln = la, nd) triple layer
  t`-type nickelates},\ }\href {https://doi.org/10.1021/ic701480v} {\bibfield
  {journal} {\bibinfo  {journal} {Inorganic Chemistry}\ }\textbf {\bibinfo
  {volume} {46}},\ \bibinfo {pages} {10887} (\bibinfo {year}
  {2007})}\BibitemShut {NoStop}%
\bibitem [{\citenamefont {Poltavets}\ \emph {et~al.}(2009)\citenamefont
  {Poltavets}, \citenamefont {Greenblatt}, \citenamefont {Fecher},\ and\
  \citenamefont {Felser}}]{Poltavets2009electronic}%
  \BibitemOpen
  \bibfield  {author} {\bibinfo {author} {\bibfnamefont {V.~V.}\ \bibnamefont
  {Poltavets}}, \bibinfo {author} {\bibfnamefont {M.}~\bibnamefont
  {Greenblatt}}, \bibinfo {author} {\bibfnamefont {G.~H.}\ \bibnamefont
  {Fecher}},\ and\ \bibinfo {author} {\bibfnamefont {C.}~\bibnamefont
  {Felser}},\ }\bibfield  {title} {\bibinfo {title} {Electronic properties,
  band structure, and fermi surface instabilities of
  ${\mathrm{ni}}^{1+}/{\mathrm{ni}}^{2+}$ nickelate
  ${\mathrm{la}}_{3}{\mathrm{ni}}_{2}{\mathrm{o}}_{6}$, isoelectronic with
  superconducting cuprates},\ }\href
  {https://doi.org/10.1103/PhysRevLett.102.046405} {\bibfield  {journal}
  {\bibinfo  {journal} {Phys. Rev. Lett.}\ }\textbf {\bibinfo {volume} {102}},\
  \bibinfo {pages} {046405} (\bibinfo {year} {2009})}\BibitemShut {NoStop}%
\bibitem [{\citenamefont {Poltavets}\ \emph {et~al.}(2010)\citenamefont
  {Poltavets}, \citenamefont {Lokshin}, \citenamefont {Nevidomskyy},
  \citenamefont {Croft}, \citenamefont {Tyson}, \citenamefont {Hadermann},
  \citenamefont {Van~Tendeloo}, \citenamefont {Egami}, \citenamefont {Kotliar},
  \citenamefont {ApRoberts-Warren}, \citenamefont {Dioguardi}, \citenamefont
  {Curro},\ and\ \citenamefont {Greenblatt}}]{Greenblatt2010bulk}%
  \BibitemOpen
  \bibfield  {author} {\bibinfo {author} {\bibfnamefont {V.~V.}\ \bibnamefont
  {Poltavets}}, \bibinfo {author} {\bibfnamefont {K.~A.}\ \bibnamefont
  {Lokshin}}, \bibinfo {author} {\bibfnamefont {A.~H.}\ \bibnamefont
  {Nevidomskyy}}, \bibinfo {author} {\bibfnamefont {M.}~\bibnamefont {Croft}},
  \bibinfo {author} {\bibfnamefont {T.~A.}\ \bibnamefont {Tyson}}, \bibinfo
  {author} {\bibfnamefont {J.}~\bibnamefont {Hadermann}}, \bibinfo {author}
  {\bibfnamefont {G.}~\bibnamefont {Van~Tendeloo}}, \bibinfo {author}
  {\bibfnamefont {T.}~\bibnamefont {Egami}}, \bibinfo {author} {\bibfnamefont
  {G.}~\bibnamefont {Kotliar}}, \bibinfo {author} {\bibfnamefont
  {N.}~\bibnamefont {ApRoberts-Warren}}, \bibinfo {author} {\bibfnamefont
  {A.~P.}\ \bibnamefont {Dioguardi}}, \bibinfo {author} {\bibfnamefont {N.~J.}\
  \bibnamefont {Curro}},\ and\ \bibinfo {author} {\bibfnamefont
  {M.}~\bibnamefont {Greenblatt}},\ }\bibfield  {title} {\bibinfo {title} {Bulk
  magnetic order in a two-dimensional ${\mathrm{ni}}^{1+}/{\mathrm{ni}}^{2+}$
  (${d}^{9}/{d}^{8}$) nickelate, isoelectronic with superconducting cuprates},\
  }\href {https://doi.org/10.1103/PhysRevLett.104.206403} {\bibfield  {journal}
  {\bibinfo  {journal} {Phys. Rev. Lett.}\ }\textbf {\bibinfo {volume} {104}},\
  \bibinfo {pages} {206403} (\bibinfo {year} {2010})}\BibitemShut {NoStop}%
\bibitem [{\citenamefont {LaBollita}\ and\ \citenamefont
  {Botana}(2021)}]{Labollita2021electronic}%
  \BibitemOpen
  \bibfield  {author} {\bibinfo {author} {\bibfnamefont {H.}~\bibnamefont
  {LaBollita}}\ and\ \bibinfo {author} {\bibfnamefont {A.~S.}\ \bibnamefont
  {Botana}},\ }\bibfield  {title} {\bibinfo {title} {Electronic structure and
  magnetic properties of higher-order layered nickelates:
  $\mathrm{La}_{n+1}\mathrm{Ni}_{n}\mathrm{O}_{2n+2} (n=4-6)$},\ }\href
  {https://doi.org/10.1103/PhysRevB.104.035148} {\bibfield  {journal} {\bibinfo
   {journal} {Phys. Rev. B}\ }\textbf {\bibinfo {volume} {104}},\ \bibinfo
  {pages} {035148} (\bibinfo {year} {2021})}\BibitemShut {NoStop}%
\bibitem [{\citenamefont {LaBollita}\ \emph {et~al.}(2022)\citenamefont
  {LaBollita}, \citenamefont {Jung},\ and\ \citenamefont
  {Botana}}]{Labollita2022manybody}%
  \BibitemOpen
  \bibfield  {author} {\bibinfo {author} {\bibfnamefont {H.}~\bibnamefont
  {LaBollita}}, \bibinfo {author} {\bibfnamefont {M.-C.}\ \bibnamefont
  {Jung}},\ and\ \bibinfo {author} {\bibfnamefont {A.~S.}\ \bibnamefont
  {Botana}},\ }\bibfield  {title} {\bibinfo {title} {Many-body electronic
  structure of ${d}^{9\ensuremath{-}\ensuremath{\delta}}$ layered nickelates},\
  }\href {https://doi.org/10.1103/PhysRevB.106.115132} {\bibfield  {journal}
  {\bibinfo  {journal} {Phys. Rev. B}\ }\textbf {\bibinfo {volume} {106}},\
  \bibinfo {pages} {115132} (\bibinfo {year} {2022})}\BibitemShut {NoStop}%
\bibitem [{\citenamefont {Botana}\ and\ \citenamefont
  {Norman}(2020)}]{botana_similarities_2020}%
  \BibitemOpen
  \bibfield  {author} {\bibinfo {author} {\bibfnamefont {A.}~\bibnamefont
  {Botana}}\ and\ \bibinfo {author} {\bibfnamefont {M.}~\bibnamefont
  {Norman}},\ }\bibfield  {title} {\bibinfo {title} {Similarities and
  differences between {LaNiO}$_2$ and {CaCuO}$_2$ and implications for
  superconductivity},\ }\href {https://doi.org/10.1103/PhysRevX.10.011024}
  {\bibfield  {journal} {\bibinfo  {journal} {Physical Review X}\ }\textbf
  {\bibinfo {volume} {10}},\ \bibinfo {pages} {011024} (\bibinfo {year}
  {2020})}\BibitemShut {NoStop}%
\bibitem [{\citenamefont {Lu}\ \emph {et~al.}(2021)\citenamefont {Lu},
  \citenamefont {Rossi}, \citenamefont {Nag}, \citenamefont {Osada},
  \citenamefont {Li}, \citenamefont {Lee}, \citenamefont {Wang}, \citenamefont
  {Garcia-Fernandez}, \citenamefont {Agrestini}, \citenamefont {Shen},
  \citenamefont {Been}, \citenamefont {Moritz}, \citenamefont {Devereaux},
  \citenamefont {Zaanen}, \citenamefont {Hwang}, \citenamefont {Zhou},\ and\
  \citenamefont {Lee}}]{lu_magnetic_2021}%
  \BibitemOpen
  \bibfield  {author} {\bibinfo {author} {\bibfnamefont {H.}~\bibnamefont
  {Lu}}, \bibinfo {author} {\bibfnamefont {M.}~\bibnamefont {Rossi}}, \bibinfo
  {author} {\bibfnamefont {A.}~\bibnamefont {Nag}}, \bibinfo {author}
  {\bibfnamefont {M.}~\bibnamefont {Osada}}, \bibinfo {author} {\bibfnamefont
  {D.~F.}\ \bibnamefont {Li}}, \bibinfo {author} {\bibfnamefont
  {K.}~\bibnamefont {Lee}}, \bibinfo {author} {\bibfnamefont {B.~Y.}\
  \bibnamefont {Wang}}, \bibinfo {author} {\bibfnamefont {M.}~\bibnamefont
  {Garcia-Fernandez}}, \bibinfo {author} {\bibfnamefont {S.}~\bibnamefont
  {Agrestini}}, \bibinfo {author} {\bibfnamefont {Z.~X.}\ \bibnamefont {Shen}},
  \bibinfo {author} {\bibfnamefont {E.~M.}\ \bibnamefont {Been}}, \bibinfo
  {author} {\bibfnamefont {B.}~\bibnamefont {Moritz}}, \bibinfo {author}
  {\bibfnamefont {T.~P.}\ \bibnamefont {Devereaux}}, \bibinfo {author}
  {\bibfnamefont {J.}~\bibnamefont {Zaanen}}, \bibinfo {author} {\bibfnamefont
  {H.~Y.}\ \bibnamefont {Hwang}}, \bibinfo {author} {\bibfnamefont {K.-J.}\
  \bibnamefont {Zhou}},\ and\ \bibinfo {author} {\bibfnamefont {W.~S.}\
  \bibnamefont {Lee}},\ }\bibfield  {title} {\bibinfo {title} {Magnetic
  excitations in infinite-layer nickelates},\ }\href
  {https://doi.org/10.1126/science.abd7726} {\bibfield  {journal} {\bibinfo
  {journal} {Science}\ }\textbf {\bibinfo {volume} {373}},\ \bibinfo {pages}
  {213} (\bibinfo {year} {2021})},\ \bibinfo {note} {publisher: American
  Association for the Advancement of Science}\BibitemShut {NoStop}%
\bibitem [{\citenamefont {Fowlie}\ \emph {et~al.}(2022)\citenamefont {Fowlie},
  \citenamefont {Hadjimichael}, \citenamefont {Martins}, \citenamefont {Li},
  \citenamefont {Osada}, \citenamefont {Wang}, \citenamefont {Lee},
  \citenamefont {Lee}, \citenamefont {Salman}, \citenamefont {Prokscha},
  \citenamefont {Triscone}, \citenamefont {Hwang},\ and\ \citenamefont
  {Suter}}]{fowlie_intrinsic_2022}%
  \BibitemOpen
  \bibfield  {author} {\bibinfo {author} {\bibfnamefont {J.}~\bibnamefont
  {Fowlie}}, \bibinfo {author} {\bibfnamefont {M.}~\bibnamefont
  {Hadjimichael}}, \bibinfo {author} {\bibfnamefont {M.~M.}\ \bibnamefont
  {Martins}}, \bibinfo {author} {\bibfnamefont {D.}~\bibnamefont {Li}},
  \bibinfo {author} {\bibfnamefont {M.}~\bibnamefont {Osada}}, \bibinfo
  {author} {\bibfnamefont {B.~Y.}\ \bibnamefont {Wang}}, \bibinfo {author}
  {\bibfnamefont {K.}~\bibnamefont {Lee}}, \bibinfo {author} {\bibfnamefont
  {Y.}~\bibnamefont {Lee}}, \bibinfo {author} {\bibfnamefont {Z.}~\bibnamefont
  {Salman}}, \bibinfo {author} {\bibfnamefont {T.}~\bibnamefont {Prokscha}},
  \bibinfo {author} {\bibfnamefont {J.-M.}\ \bibnamefont {Triscone}}, \bibinfo
  {author} {\bibfnamefont {H.~Y.}\ \bibnamefont {Hwang}},\ and\ \bibinfo
  {author} {\bibfnamefont {A.}~\bibnamefont {Suter}},\ }\bibfield  {title}
  {\bibinfo {title} {Intrinsic magnetism in superconducting infinite-layer
  nickelates},\ }\href {https://doi.org/10.1038/s41567-022-01684-y} {\bibfield
  {journal} {\bibinfo  {journal} {Nature Physics}\ }\textbf {\bibinfo {volume}
  {18}},\ \bibinfo {pages} {1043} (\bibinfo {year} {2022})},\ \bibinfo {note}
  {arXiv:2201.11943}\BibitemShut {NoStop}%
\bibitem [{\citenamefont {Rossi}\ \emph {et~al.}(2022)\citenamefont {Rossi},
  \citenamefont {Osada}, \citenamefont {Choi}, \citenamefont {Agrestini},
  \citenamefont {Jost}, \citenamefont {Lee}, \citenamefont {Lu}, \citenamefont
  {Wang}, \citenamefont {Lee}, \citenamefont {Nag}, \citenamefont {Chuang},
  \citenamefont {Kuo}, \citenamefont {Lee}, \citenamefont {Moritz},
  \citenamefont {Devereaux}, \citenamefont {Shen}, \citenamefont {Lee},
  \citenamefont {Zhou}, \citenamefont {Hwang},\ and\ \citenamefont
  {Lee}}]{rossi_broken_2022}%
  \BibitemOpen
  \bibfield  {author} {\bibinfo {author} {\bibfnamefont {M.}~\bibnamefont
  {Rossi}}, \bibinfo {author} {\bibfnamefont {M.}~\bibnamefont {Osada}},
  \bibinfo {author} {\bibfnamefont {J.}~\bibnamefont {Choi}}, \bibinfo {author}
  {\bibfnamefont {S.}~\bibnamefont {Agrestini}}, \bibinfo {author}
  {\bibfnamefont {D.}~\bibnamefont {Jost}}, \bibinfo {author} {\bibfnamefont
  {Y.}~\bibnamefont {Lee}}, \bibinfo {author} {\bibfnamefont {H.}~\bibnamefont
  {Lu}}, \bibinfo {author} {\bibfnamefont {B.~Y.}\ \bibnamefont {Wang}},
  \bibinfo {author} {\bibfnamefont {K.}~\bibnamefont {Lee}}, \bibinfo {author}
  {\bibfnamefont {A.}~\bibnamefont {Nag}}, \bibinfo {author} {\bibfnamefont
  {Y.-D.}\ \bibnamefont {Chuang}}, \bibinfo {author} {\bibfnamefont {C.-T.}\
  \bibnamefont {Kuo}}, \bibinfo {author} {\bibfnamefont {S.-J.}\ \bibnamefont
  {Lee}}, \bibinfo {author} {\bibfnamefont {B.}~\bibnamefont {Moritz}},
  \bibinfo {author} {\bibfnamefont {T.~P.}\ \bibnamefont {Devereaux}}, \bibinfo
  {author} {\bibfnamefont {Z.-X.}\ \bibnamefont {Shen}}, \bibinfo {author}
  {\bibfnamefont {J.-S.}\ \bibnamefont {Lee}}, \bibinfo {author} {\bibfnamefont
  {K.-J.}\ \bibnamefont {Zhou}}, \bibinfo {author} {\bibfnamefont {H.~Y.}\
  \bibnamefont {Hwang}},\ and\ \bibinfo {author} {\bibfnamefont {W.-S.}\
  \bibnamefont {Lee}},\ }\bibfield  {title} {\bibinfo {title} {A broken
  translational symmetry state in an infinite-layer nickelate},\ }\href
  {https://doi.org/10.1038/s41567-022-01660-6} {\bibfield  {journal} {\bibinfo
  {journal} {Nature Physics}\ }\textbf {\bibinfo {volume} {18}},\ \bibinfo
  {pages} {869} (\bibinfo {year} {2022})}\BibitemShut {NoStop}%
\bibitem [{\citenamefont {Tam}\ \emph {et~al.}(2022)\citenamefont {Tam},
  \citenamefont {Choi}, \citenamefont {Ding}, \citenamefont {Agrestini},
  \citenamefont {Nag}, \citenamefont {Wu}, \citenamefont {Huang}, \citenamefont
  {Luo}, \citenamefont {Gao}, \citenamefont {García-Fernández}, \citenamefont
  {Qiao},\ and\ \citenamefont {Zhou}}]{tam_charge_2022}%
  \BibitemOpen
  \bibfield  {author} {\bibinfo {author} {\bibfnamefont {C.~C.}\ \bibnamefont
  {Tam}}, \bibinfo {author} {\bibfnamefont {J.}~\bibnamefont {Choi}}, \bibinfo
  {author} {\bibfnamefont {X.}~\bibnamefont {Ding}}, \bibinfo {author}
  {\bibfnamefont {S.}~\bibnamefont {Agrestini}}, \bibinfo {author}
  {\bibfnamefont {A.}~\bibnamefont {Nag}}, \bibinfo {author} {\bibfnamefont
  {M.}~\bibnamefont {Wu}}, \bibinfo {author} {\bibfnamefont {B.}~\bibnamefont
  {Huang}}, \bibinfo {author} {\bibfnamefont {H.}~\bibnamefont {Luo}}, \bibinfo
  {author} {\bibfnamefont {P.}~\bibnamefont {Gao}}, \bibinfo {author}
  {\bibfnamefont {M.}~\bibnamefont {García-Fernández}}, \bibinfo {author}
  {\bibfnamefont {L.}~\bibnamefont {Qiao}},\ and\ \bibinfo {author}
  {\bibfnamefont {K.-J.}\ \bibnamefont {Zhou}},\ }\bibfield  {title} {\bibinfo
  {title} {Charge density waves in infinite-layer {NdNiO}$_2$ nickelates},\
  }\href {https://doi.org/10.1038/s41563-022-01330-1} {\bibfield  {journal}
  {\bibinfo  {journal} {Nature Materials}\ }\textbf {\bibinfo {volume} {21}},\
  \bibinfo {pages} {1116} (\bibinfo {year} {2022})}\BibitemShut {NoStop}%
\bibitem [{\citenamefont {Lee}\ \emph {et~al.}(2023)\citenamefont {Lee},
  \citenamefont {Wang}, \citenamefont {Osada}, \citenamefont {Goodge},
  \citenamefont {Wang}, \citenamefont {Lee}, \citenamefont {Harvey},
  \citenamefont {Kim}, \citenamefont {Yu}, \citenamefont {Murthy},
  \citenamefont {Raghu}, \citenamefont {Kourkoutis},\ and\ \citenamefont
  {Hwang}}]{lee_linear_character_2023}%
  \BibitemOpen
  \bibfield  {author} {\bibinfo {author} {\bibfnamefont {K.}~\bibnamefont
  {Lee}}, \bibinfo {author} {\bibfnamefont {B.~Y.}\ \bibnamefont {Wang}},
  \bibinfo {author} {\bibfnamefont {M.}~\bibnamefont {Osada}}, \bibinfo
  {author} {\bibfnamefont {B.~H.}\ \bibnamefont {Goodge}}, \bibinfo {author}
  {\bibfnamefont {T.~C.}\ \bibnamefont {Wang}}, \bibinfo {author}
  {\bibfnamefont {Y.}~\bibnamefont {Lee}}, \bibinfo {author} {\bibfnamefont
  {S.}~\bibnamefont {Harvey}}, \bibinfo {author} {\bibfnamefont {W.~J.}\
  \bibnamefont {Kim}}, \bibinfo {author} {\bibfnamefont {Y.}~\bibnamefont
  {Yu}}, \bibinfo {author} {\bibfnamefont {C.}~\bibnamefont {Murthy}}, \bibinfo
  {author} {\bibfnamefont {S.}~\bibnamefont {Raghu}}, \bibinfo {author}
  {\bibfnamefont {L.~F.}\ \bibnamefont {Kourkoutis}},\ and\ \bibinfo {author}
  {\bibfnamefont {H.~Y.}\ \bibnamefont {Hwang}},\ }\bibfield  {title} {\bibinfo
  {title} {Linear-in-temperature resistivity for optimally superconducting
  ({{Nd}},{{Sr}}){{NiO2}}},\ }\href
  {https://doi.org/10.1038/s41586-023-06129-x} {\bibfield  {journal} {\bibinfo
  {journal} {Nature}\ }\textbf {\bibinfo {volume} {619}},\ \bibinfo {pages}
  {288} (\bibinfo {year} {2023})}\BibitemShut {NoStop}%
\bibitem [{\citenamefont {Chen}\ \emph {et~al.}(2022)\citenamefont {Chen},
  \citenamefont {Osada}, \citenamefont {Li}, \citenamefont {Been},
  \citenamefont {Chen}, \citenamefont {Hashimoto}, \citenamefont {Lu},
  \citenamefont {Mo}, \citenamefont {Lee}, \citenamefont {Wang}, \citenamefont
  {Rodolakis}, \citenamefont {McChesney}, \citenamefont {Jia}, \citenamefont
  {Moritz}, \citenamefont {Devereaux}, \citenamefont {Hwang},\ and\
  \citenamefont {Shen}}]{chen_electronic_2022}%
  \BibitemOpen
  \bibfield  {author} {\bibinfo {author} {\bibfnamefont {Z.}~\bibnamefont
  {Chen}}, \bibinfo {author} {\bibfnamefont {M.}~\bibnamefont {Osada}},
  \bibinfo {author} {\bibfnamefont {D.}~\bibnamefont {Li}}, \bibinfo {author}
  {\bibfnamefont {E.~M.}\ \bibnamefont {Been}}, \bibinfo {author}
  {\bibfnamefont {S.-D.}\ \bibnamefont {Chen}}, \bibinfo {author}
  {\bibfnamefont {M.}~\bibnamefont {Hashimoto}}, \bibinfo {author}
  {\bibfnamefont {D.}~\bibnamefont {Lu}}, \bibinfo {author} {\bibfnamefont
  {S.-K.}\ \bibnamefont {Mo}}, \bibinfo {author} {\bibfnamefont
  {K.}~\bibnamefont {Lee}}, \bibinfo {author} {\bibfnamefont {B.~Y.}\
  \bibnamefont {Wang}}, \bibinfo {author} {\bibfnamefont {F.}~\bibnamefont
  {Rodolakis}}, \bibinfo {author} {\bibfnamefont {J.~L.}\ \bibnamefont
  {McChesney}}, \bibinfo {author} {\bibfnamefont {C.}~\bibnamefont {Jia}},
  \bibinfo {author} {\bibfnamefont {B.}~\bibnamefont {Moritz}}, \bibinfo
  {author} {\bibfnamefont {T.~P.}\ \bibnamefont {Devereaux}}, \bibinfo {author}
  {\bibfnamefont {H.~Y.}\ \bibnamefont {Hwang}},\ and\ \bibinfo {author}
  {\bibfnamefont {Z.-X.}\ \bibnamefont {Shen}},\ }\bibfield  {title} {\bibinfo
  {title} {Electronic structure of superconducting nickelates probed by
  resonant photoemission spectroscopy},\ }\href
  {https://doi.org/10.1016/j.matt.2022.01.020} {\bibfield  {journal} {\bibinfo
  {journal} {Matter}\ }\textbf {\bibinfo {volume} {5}},\ \bibinfo {pages}
  {1806} (\bibinfo {year} {2022})}\BibitemShut {NoStop}%
\bibitem [{\citenamefont {Li}\ \emph {et~al.}(2022)\citenamefont {Li},
  \citenamefont {Hao}, \citenamefont {Zhang}, \citenamefont {Gordon},
  \citenamefont {Linn}, \citenamefont {Zheng}, \citenamefont {Zhou},
  \citenamefont {Mitchell},\ and\ \citenamefont {Dessau}}]{li_electronic_2022}%
  \BibitemOpen
  \bibfield  {author} {\bibinfo {author} {\bibfnamefont {H.}~\bibnamefont
  {Li}}, \bibinfo {author} {\bibfnamefont {P.}~\bibnamefont {Hao}}, \bibinfo
  {author} {\bibfnamefont {J.}~\bibnamefont {Zhang}}, \bibinfo {author}
  {\bibfnamefont {K.}~\bibnamefont {Gordon}}, \bibinfo {author} {\bibfnamefont
  {A.~G.}\ \bibnamefont {Linn}}, \bibinfo {author} {\bibfnamefont
  {H.}~\bibnamefont {Zheng}}, \bibinfo {author} {\bibfnamefont
  {X.}~\bibnamefont {Zhou}}, \bibinfo {author} {\bibfnamefont {J.~F.}\
  \bibnamefont {Mitchell}},\ and\ \bibinfo {author} {\bibfnamefont {D.~S.}\
  \bibnamefont {Dessau}},\ }\bibfield  {title} {\bibinfo {title} {Electronic
  structure and correlations in planar trilayer nickelate
  {Pr}$_4${Ni}$_3${O}$_8$},\ }\bibfield  {journal} {\bibinfo  {journal}
  {arXiv.2207.13633}\ }\href {https://doi.org/10.48550/arXiv.2207.13633}
  {10.48550/arXiv.2207.13633} (\bibinfo {year} {2022})\BibitemShut {NoStop}%
\bibitem [{\citenamefont {Gourgout}\ \emph {et~al.}(2022)\citenamefont
  {Gourgout}, \citenamefont {Grissonnanche}, \citenamefont {Lalibert\'{e}},
  \citenamefont {Ataei}, \citenamefont {Chen}, \citenamefont {Verret},
  \citenamefont {Zhou}, \citenamefont {Mravlje}, \citenamefont {Georges},
  \citenamefont {Doiron-Leyraud},\ and\ \citenamefont
  {Taillefer}}]{Gourgout2022Seebeck}%
  \BibitemOpen
  \bibfield  {author} {\bibinfo {author} {\bibfnamefont {A.}~\bibnamefont
  {Gourgout}}, \bibinfo {author} {\bibfnamefont {G.}~\bibnamefont
  {Grissonnanche}}, \bibinfo {author} {\bibfnamefont {F.}~\bibnamefont
  {Lalibert\'{e}}}, \bibinfo {author} {\bibfnamefont {A.}~\bibnamefont
  {Ataei}}, \bibinfo {author} {\bibfnamefont {L.}~\bibnamefont {Chen}},
  \bibinfo {author} {\bibfnamefont {S.}~\bibnamefont {Verret}}, \bibinfo
  {author} {\bibfnamefont {J.-S.}\ \bibnamefont {Zhou}}, \bibinfo {author}
  {\bibfnamefont {J.}~\bibnamefont {Mravlje}}, \bibinfo {author} {\bibfnamefont
  {A.}~\bibnamefont {Georges}}, \bibinfo {author} {\bibfnamefont
  {N.}~\bibnamefont {Doiron-Leyraud}},\ and\ \bibinfo {author} {\bibfnamefont
  {L.}~\bibnamefont {Taillefer}},\ }\bibfield  {title} {\bibinfo {title}
  {Seebeck coefficient in a cuprate superconductor: Particle-hole asymmetry in
  the strange metal phase and fermi surface transformation in the pseudogap
  phase},\ }\href {https://doi.org/10.1103/PhysRevX.12.011037} {\bibfield
  {journal} {\bibinfo  {journal} {Physical Review X}\ }\textbf {\bibinfo
  {volume} {12}},\ \bibinfo {pages} {011037} (\bibinfo {year}
  {2022})}\BibitemShut {NoStop}%
\bibitem [{\citenamefont {Kondo}\ \emph {et~al.}(2005)\citenamefont {Kondo},
  \citenamefont {Takeuchi}, \citenamefont {Mizutani}, \citenamefont {Yokoya},
  \citenamefont {Tsuda},\ and\ \citenamefont {Shin}}]{Kondo2005Contribution}%
  \BibitemOpen
  \bibfield  {author} {\bibinfo {author} {\bibfnamefont {T.}~\bibnamefont
  {Kondo}}, \bibinfo {author} {\bibfnamefont {T.}~\bibnamefont {Takeuchi}},
  \bibinfo {author} {\bibfnamefont {U.}~\bibnamefont {Mizutani}}, \bibinfo
  {author} {\bibfnamefont {T.}~\bibnamefont {Yokoya}}, \bibinfo {author}
  {\bibfnamefont {S.}~\bibnamefont {Tsuda}},\ and\ \bibinfo {author}
  {\bibfnamefont {S.}~\bibnamefont {Shin}},\ }\bibfield  {title} {\bibinfo
  {title} {Contribution of electronic structure to thermoelectric power in
  ({Bi,Pb})$_2$({Sr,La} )$_2${CuO}$_{\rm 6 + \delta}$},\ }\href
  {https://doi.org/10.1103/PhysRevB.72.024533} {\bibfield  {journal} {\bibinfo
  {journal} {Physical Review B}\ }\textbf {\bibinfo {volume} {72}},\ \bibinfo
  {pages} {024533} (\bibinfo {year} {2005})}\BibitemShut {NoStop}%
\bibitem [{\citenamefont {Cooper}\ \emph {et~al.}(2009)\citenamefont {Cooper},
  \citenamefont {Wang}, \citenamefont {Vignolle}, \citenamefont {Lipscombe},
  \citenamefont {Hayden}, \citenamefont {Tanabe}, \citenamefont {Adachi},
  \citenamefont {Koike}, \citenamefont {Nohara}, \citenamefont {Takagi},
  \citenamefont {Proust},\ and\ \citenamefont {Hussey}}]{Cooper2009Anomalous}%
  \BibitemOpen
  \bibfield  {author} {\bibinfo {author} {\bibfnamefont {R.~A.}\ \bibnamefont
  {Cooper}}, \bibinfo {author} {\bibfnamefont {Y.}~\bibnamefont {Wang}},
  \bibinfo {author} {\bibfnamefont {B.}~\bibnamefont {Vignolle}}, \bibinfo
  {author} {\bibfnamefont {O.~J.}\ \bibnamefont {Lipscombe}}, \bibinfo {author}
  {\bibfnamefont {S.~M.}\ \bibnamefont {Hayden}}, \bibinfo {author}
  {\bibfnamefont {Y.}~\bibnamefont {Tanabe}}, \bibinfo {author} {\bibfnamefont
  {T.}~\bibnamefont {Adachi}}, \bibinfo {author} {\bibfnamefont
  {Y.}~\bibnamefont {Koike}}, \bibinfo {author} {\bibfnamefont
  {M.}~\bibnamefont {Nohara}}, \bibinfo {author} {\bibfnamefont
  {H.}~\bibnamefont {Takagi}}, \bibinfo {author} {\bibfnamefont
  {C.}~\bibnamefont {Proust}},\ and\ \bibinfo {author} {\bibfnamefont {N.~E.}\
  \bibnamefont {Hussey}},\ }\bibfield  {title} {\bibinfo {title} {Anomalous
  criticality in the electrical resistivity of la2-xsrxcuo4},\ }\href
  {https://doi.org/10.1126/science.1165015} {\bibfield  {journal} {\bibinfo
  {journal} {Science}\ }\textbf {\bibinfo {volume} {323}},\ \bibinfo {pages}
  {603} (\bibinfo {year} {2009})}\BibitemShut {NoStop}%
\bibitem [{\citenamefont {Berben}\ \emph {et~al.}(2022)\citenamefont {Berben},
  \citenamefont {Smit}, \citenamefont {Duffy}, \citenamefont {Hsu},
  \citenamefont {Bawden}, \citenamefont {Heringa}, \citenamefont {Gerritsen},
  \citenamefont {Cassanelli}, \citenamefont {Feng}, \citenamefont {Bron},
  \citenamefont {Van~Heumen}, \citenamefont {Huang}, \citenamefont {Bertran},
  \citenamefont {Kim}, \citenamefont {Cacho}, \citenamefont {Carrington},
  \citenamefont {Golden},\ and\ \citenamefont {Hussey}}]{berben_2022a}%
  \BibitemOpen
  \bibfield  {author} {\bibinfo {author} {\bibfnamefont {M.}~\bibnamefont
  {Berben}}, \bibinfo {author} {\bibfnamefont {S.}~\bibnamefont {Smit}},
  \bibinfo {author} {\bibfnamefont {C.}~\bibnamefont {Duffy}}, \bibinfo
  {author} {\bibfnamefont {Y.-T.}\ \bibnamefont {Hsu}}, \bibinfo {author}
  {\bibfnamefont {L.}~\bibnamefont {Bawden}}, \bibinfo {author} {\bibfnamefont
  {F.}~\bibnamefont {Heringa}}, \bibinfo {author} {\bibfnamefont
  {F.}~\bibnamefont {Gerritsen}}, \bibinfo {author} {\bibfnamefont
  {S.}~\bibnamefont {Cassanelli}}, \bibinfo {author} {\bibfnamefont
  {X.}~\bibnamefont {Feng}}, \bibinfo {author} {\bibfnamefont {S.}~\bibnamefont
  {Bron}}, \bibinfo {author} {\bibfnamefont {E.}~\bibnamefont {Van~Heumen}},
  \bibinfo {author} {\bibfnamefont {Y.}~\bibnamefont {Huang}}, \bibinfo
  {author} {\bibfnamefont {F.}~\bibnamefont {Bertran}}, \bibinfo {author}
  {\bibfnamefont {T.~K.}\ \bibnamefont {Kim}}, \bibinfo {author} {\bibfnamefont
  {C.}~\bibnamefont {Cacho}}, \bibinfo {author} {\bibfnamefont
  {A.}~\bibnamefont {Carrington}}, \bibinfo {author} {\bibfnamefont {M.~S.}\
  \bibnamefont {Golden}},\ and\ \bibinfo {author} {\bibfnamefont {N.~E.}\
  \bibnamefont {Hussey}},\ }\bibfield  {title} {\bibinfo {title}
  {Superconducting dome and pseudogap endpoint in {{Bi2201}}},\ }\href
  {https://doi.org/10.1103/PhysRevMaterials.6.044804} {\bibfield  {journal}
  {\bibinfo  {journal} {Physical Review Materials}\ }\textbf {\bibinfo {volume}
  {6}},\ \bibinfo {pages} {044804} (\bibinfo {year} {2022})}\BibitemShut
  {NoStop}%
\bibitem [{\citenamefont {Pan}\ \emph {et~al.}(2022{\natexlab{b}})\citenamefont
  {Pan}, \citenamefont {Song}, \citenamefont {Ferenc~Segedin}, \citenamefont
  {Jung}, \citenamefont {El-Sherif}, \citenamefont {Fleck}, \citenamefont
  {Goodge}, \citenamefont {Doyle}, \citenamefont {C\'ordova~Carrizales},
  \citenamefont {N'Diaye}, \citenamefont {Shafer}, \citenamefont {Paik},
  \citenamefont {Kourkoutis}, \citenamefont {El~Baggari}, \citenamefont
  {Botana}, \citenamefont {Brooks},\ and\ \citenamefont
  {Mundy}}]{Pan2022Synthesis}%
  \BibitemOpen
  \bibfield  {author} {\bibinfo {author} {\bibfnamefont {G.~A.}\ \bibnamefont
  {Pan}}, \bibinfo {author} {\bibfnamefont {Q.}~\bibnamefont {Song}}, \bibinfo
  {author} {\bibfnamefont {D.}~\bibnamefont {Ferenc~Segedin}}, \bibinfo
  {author} {\bibfnamefont {M.-C.}\ \bibnamefont {Jung}}, \bibinfo {author}
  {\bibfnamefont {H.}~\bibnamefont {El-Sherif}}, \bibinfo {author}
  {\bibfnamefont {E.~E.}\ \bibnamefont {Fleck}}, \bibinfo {author}
  {\bibfnamefont {B.~H.}\ \bibnamefont {Goodge}}, \bibinfo {author}
  {\bibfnamefont {S.}~\bibnamefont {Doyle}}, \bibinfo {author} {\bibfnamefont
  {D.}~\bibnamefont {C\'ordova~Carrizales}}, \bibinfo {author} {\bibfnamefont
  {A.~T.}\ \bibnamefont {N'Diaye}}, \bibinfo {author} {\bibfnamefont
  {P.}~\bibnamefont {Shafer}}, \bibinfo {author} {\bibfnamefont
  {H.}~\bibnamefont {Paik}}, \bibinfo {author} {\bibfnamefont {L.~F.}\
  \bibnamefont {Kourkoutis}}, \bibinfo {author} {\bibfnamefont
  {I.}~\bibnamefont {El~Baggari}}, \bibinfo {author} {\bibfnamefont {A.~S.}\
  \bibnamefont {Botana}}, \bibinfo {author} {\bibfnamefont {C.~M.}\
  \bibnamefont {Brooks}},\ and\ \bibinfo {author} {\bibfnamefont {J.~A.}\
  \bibnamefont {Mundy}},\ }\bibfield  {title} {\bibinfo {title} {Synthesis and
  electronic properties of {Nd}$_{\rm n+1}${Ni}$_{\rm n}${O}$_{\rm 3n+1}$
  {Ruddlesden-Popper} nickelate thin films},\ }\href
  {https://doi.org/10.1103/PhysRevMaterials.6.055003} {\bibfield  {journal}
  {\bibinfo  {journal} {Phys. Rev. Materials}\ }\textbf {\bibinfo {volume}
  {6}},\ \bibinfo {pages} {055003} (\bibinfo {year}
  {2022}{\natexlab{b}})}\BibitemShut {NoStop}%
\bibitem [{\citenamefont {Schnelle}\ \emph {et~al.}(2001)\citenamefont
  {Schnelle}, \citenamefont {Fischer},\ and\ \citenamefont
  {Gmelin}}]{Schnelle_2001}%
  \BibitemOpen
  \bibfield  {author} {\bibinfo {author} {\bibfnamefont {W.}~\bibnamefont
  {Schnelle}}, \bibinfo {author} {\bibfnamefont {R.}~\bibnamefont {Fischer}},\
  and\ \bibinfo {author} {\bibfnamefont {E.}~\bibnamefont {Gmelin}},\
  }\bibfield  {title} {\bibinfo {title} {Specific heat capacity and thermal
  conductivity of {NdGaO}$_3$ and {LaAlO}$_3$ single crystals at low
  temperatures},\ }\href {https://doi.org/10.1088/0022-3727/34/6/302}
  {\bibfield  {journal} {\bibinfo  {journal} {Journal of Physics D: Applied
  Physics}\ }\textbf {\bibinfo {volume} {34}},\ \bibinfo {pages} {846}
  (\bibinfo {year} {2001})}\BibitemShut {NoStop}%
\bibitem [{\citenamefont {Ding}\ \emph {et~al.}(2022)\citenamefont {Ding},
  \citenamefont {Shen}, \citenamefont {Leng}, \citenamefont {Xu}, \citenamefont
  {Zhao}, \citenamefont {Zhao}, \citenamefont {Sui}, \citenamefont {Wu},
  \citenamefont {Xiao}, \citenamefont {Zu}, \citenamefont {Huang},
  \citenamefont {Luo}, \citenamefont {Yu},\ and\ \citenamefont
  {Qiao}}]{ding_stability_2022}%
  \BibitemOpen
  \bibfield  {author} {\bibinfo {author} {\bibfnamefont {X.}~\bibnamefont
  {Ding}}, \bibinfo {author} {\bibfnamefont {S.}~\bibnamefont {Shen}}, \bibinfo
  {author} {\bibfnamefont {H.}~\bibnamefont {Leng}}, \bibinfo {author}
  {\bibfnamefont {M.}~\bibnamefont {Xu}}, \bibinfo {author} {\bibfnamefont
  {Y.}~\bibnamefont {Zhao}}, \bibinfo {author} {\bibfnamefont {J.}~\bibnamefont
  {Zhao}}, \bibinfo {author} {\bibfnamefont {X.}~\bibnamefont {Sui}}, \bibinfo
  {author} {\bibfnamefont {X.}~\bibnamefont {Wu}}, \bibinfo {author}
  {\bibfnamefont {H.}~\bibnamefont {Xiao}}, \bibinfo {author} {\bibfnamefont
  {X.}~\bibnamefont {Zu}}, \bibinfo {author} {\bibfnamefont {B.}~\bibnamefont
  {Huang}}, \bibinfo {author} {\bibfnamefont {H.}~\bibnamefont {Luo}}, \bibinfo
  {author} {\bibfnamefont {P.}~\bibnamefont {Yu}},\ and\ \bibinfo {author}
  {\bibfnamefont {L.}~\bibnamefont {Qiao}},\ }\bibfield  {title} {\bibinfo
  {title} {Stability of superconducting {Nd}$_{0.8}${Sr}$_{0.2}${NiO}$_2$ thin
  films},\ }\href {https://doi.org/10.1007/s11433-021-1871-x} {\bibfield
  {journal} {\bibinfo  {journal} {Science China Physics, Mechanics \&
  Astronomy}\ }\textbf {\bibinfo {volume} {65}},\ \bibinfo {pages} {267411}
  (\bibinfo {year} {2022})}\BibitemShut {NoStop}%
\bibitem [{\citenamefont {Kresse}\ and\ \citenamefont
  {Furthm\"uller}(1996)}]{Kresse1996}%
  \BibitemOpen
  \bibfield  {author} {\bibinfo {author} {\bibfnamefont {G.}~\bibnamefont
  {Kresse}}\ and\ \bibinfo {author} {\bibfnamefont {J.}~\bibnamefont
  {Furthm\"uller}},\ }\bibfield  {title} {\bibinfo {title} {Efficient iterative
  schemes for ab initio total-energy calculations using a plane-wave basis
  set},\ }\href {https://doi.org/10.1103/PhysRevB.54.11169} {\bibfield
  {journal} {\bibinfo  {journal} {Phys. Rev. B}\ }\textbf {\bibinfo {volume}
  {54}},\ \bibinfo {pages} {11169} (\bibinfo {year} {1996})}\BibitemShut
  {NoStop}%
\bibitem [{\citenamefont {Grissonnanche}\ \emph {et~al.}(2021)\citenamefont
  {Grissonnanche}, \citenamefont {Fang}, \citenamefont {Legros}, \citenamefont
  {Verret}, \citenamefont {Laliberté}, \citenamefont {Collignon},
  \citenamefont {Zhou}, \citenamefont {Graf}, \citenamefont {Goddard},
  \citenamefont {Taillefer},\ and\ \citenamefont
  {Ramshaw}}]{Grissonnanche2021LinearIn}%
  \BibitemOpen
  \bibfield  {author} {\bibinfo {author} {\bibfnamefont {G.}~\bibnamefont
  {Grissonnanche}}, \bibinfo {author} {\bibfnamefont {Y.}~\bibnamefont {Fang}},
  \bibinfo {author} {\bibfnamefont {A.}~\bibnamefont {Legros}}, \bibinfo
  {author} {\bibfnamefont {S.}~\bibnamefont {Verret}}, \bibinfo {author}
  {\bibfnamefont {F.}~\bibnamefont {Laliberté}}, \bibinfo {author}
  {\bibfnamefont {C.}~\bibnamefont {Collignon}}, \bibinfo {author}
  {\bibfnamefont {J.}~\bibnamefont {Zhou}}, \bibinfo {author} {\bibfnamefont
  {D.}~\bibnamefont {Graf}}, \bibinfo {author} {\bibfnamefont {P.~A.}\
  \bibnamefont {Goddard}}, \bibinfo {author} {\bibfnamefont {L.}~\bibnamefont
  {Taillefer}},\ and\ \bibinfo {author} {\bibfnamefont {B.~J.}\ \bibnamefont
  {Ramshaw}},\ }\bibfield  {title} {\bibinfo {title} {Linear-in temperature
  resistivity from an isotropic planckian scattering rate},\ }\href
  {https://doi.org/10.1038/s41586-021-03697-8} {\bibfield  {journal} {\bibinfo
  {journal} {Nature}\ }\textbf {\bibinfo {volume} {595}},\ \bibinfo {pages}
  {667} (\bibinfo {year} {2021})}\BibitemShut {NoStop}%
\bibitem [{\citenamefont {Fang}\ \emph {et~al.}(2022)\citenamefont {Fang},
  \citenamefont {Grissonnanche}, \citenamefont {Legros}, \citenamefont
  {Verret}, \citenamefont {Laliberté}, \citenamefont {Collignon},
  \citenamefont {Ataei}, \citenamefont {Dion}, \citenamefont {Zhou},
  \citenamefont {Graf}, \citenamefont {Lawler}, \citenamefont {Goddard},
  \citenamefont {Taillefer},\ and\ \citenamefont {Ramshaw}}]{Fang2022Fermi}%
  \BibitemOpen
  \bibfield  {author} {\bibinfo {author} {\bibfnamefont {Y.}~\bibnamefont
  {Fang}}, \bibinfo {author} {\bibfnamefont {G.}~\bibnamefont {Grissonnanche}},
  \bibinfo {author} {\bibfnamefont {A.}~\bibnamefont {Legros}}, \bibinfo
  {author} {\bibfnamefont {S.}~\bibnamefont {Verret}}, \bibinfo {author}
  {\bibfnamefont {F.}~\bibnamefont {Laliberté}}, \bibinfo {author}
  {\bibfnamefont {C.}~\bibnamefont {Collignon}}, \bibinfo {author}
  {\bibfnamefont {A.}~\bibnamefont {Ataei}}, \bibinfo {author} {\bibfnamefont
  {M.}~\bibnamefont {Dion}}, \bibinfo {author} {\bibfnamefont {J.}~\bibnamefont
  {Zhou}}, \bibinfo {author} {\bibfnamefont {D.}~\bibnamefont {Graf}}, \bibinfo
  {author} {\bibfnamefont {M.~J.}\ \bibnamefont {Lawler}}, \bibinfo {author}
  {\bibfnamefont {P.~A.}\ \bibnamefont {Goddard}}, \bibinfo {author}
  {\bibfnamefont {L.}~\bibnamefont {Taillefer}},\ and\ \bibinfo {author}
  {\bibfnamefont {B.~J.}\ \bibnamefont {Ramshaw}},\ }\bibfield  {title}
  {\bibinfo {title} {Fermi surface transformation at the pseudogap critical
  point of a cuprate superconductor},\ }\href
  {https://doi.org/10.1038/s41567-022-01514-1} {\bibfield  {journal} {\bibinfo
  {journal} {Nature Physics}\ }\textbf {\bibinfo {volume} {18}},\ \bibinfo
  {pages} {558} (\bibinfo {year} {2022})}\BibitemShut {NoStop}%
\bibitem [{\citenamefont {Ataei}\ \emph {et~al.}(2022)\citenamefont {Ataei},
  \citenamefont {Gourgout}, \citenamefont {Grissonnanche}, \citenamefont
  {Chen}, \citenamefont {Baglo}, \citenamefont {Boulanger}, \citenamefont
  {Laliberté}, \citenamefont {Badoux}, \citenamefont {Doiron-Leyraud},
  \citenamefont {Oliviero}, \citenamefont {Benhabib}, \citenamefont
  {Vignolles}, \citenamefont {Zhou}, \citenamefont {Ono}, \citenamefont
  {Takagi}, \citenamefont {Proust},\ and\ \citenamefont
  {Taillefer}}]{Ataei2022Electrons}%
  \BibitemOpen
  \bibfield  {author} {\bibinfo {author} {\bibfnamefont {A.}~\bibnamefont
  {Ataei}}, \bibinfo {author} {\bibfnamefont {A.}~\bibnamefont {Gourgout}},
  \bibinfo {author} {\bibfnamefont {G.}~\bibnamefont {Grissonnanche}}, \bibinfo
  {author} {\bibfnamefont {L.}~\bibnamefont {Chen}}, \bibinfo {author}
  {\bibfnamefont {J.}~\bibnamefont {Baglo}}, \bibinfo {author} {\bibfnamefont
  {M.-E.}\ \bibnamefont {Boulanger}}, \bibinfo {author} {\bibfnamefont
  {F.}~\bibnamefont {Laliberté}}, \bibinfo {author} {\bibfnamefont
  {S.}~\bibnamefont {Badoux}}, \bibinfo {author} {\bibfnamefont
  {N.}~\bibnamefont {Doiron-Leyraud}}, \bibinfo {author} {\bibfnamefont
  {V.}~\bibnamefont {Oliviero}}, \bibinfo {author} {\bibfnamefont
  {S.}~\bibnamefont {Benhabib}}, \bibinfo {author} {\bibfnamefont
  {D.}~\bibnamefont {Vignolles}}, \bibinfo {author} {\bibfnamefont {J.-S.}\
  \bibnamefont {Zhou}}, \bibinfo {author} {\bibfnamefont {S.}~\bibnamefont
  {Ono}}, \bibinfo {author} {\bibfnamefont {H.}~\bibnamefont {Takagi}},
  \bibinfo {author} {\bibfnamefont {C.}~\bibnamefont {Proust}},\ and\ \bibinfo
  {author} {\bibfnamefont {L.}~\bibnamefont {Taillefer}},\ }\bibfield  {title}
  {\bibinfo {title} {Electrons with {Planckian} scattering obey standard
  orbital motion in a magnetic field},\ }\href
  {https://doi.org/10.1038/s41567-022-01763-0} {\bibfield  {journal} {\bibinfo
  {journal} {Nature Physics}\ ,\ \bibinfo {pages} {1}} (\bibinfo {year}
  {2022})},\ \bibinfo {note} {publisher: Nature Publishing Group}\BibitemShut
  {NoStop}%
\bibitem [{\citenamefont {Cheng}\ \emph {et~al.}(2012)\citenamefont {Cheng},
  \citenamefont {Zhou}, \citenamefont {Goodenough}, \citenamefont {Zhou},
  \citenamefont {Matsubayashi}, \citenamefont {Uwatoko}, \citenamefont {Kong},
  \citenamefont {Jin}, \citenamefont {Yang},\ and\ \citenamefont
  {Shen}}]{cheng2012pressure}%
  \BibitemOpen
  \bibfield  {author} {\bibinfo {author} {\bibfnamefont {J.-G.}\ \bibnamefont
  {Cheng}}, \bibinfo {author} {\bibfnamefont {J.-S.}\ \bibnamefont {Zhou}},
  \bibinfo {author} {\bibfnamefont {J.~B.}\ \bibnamefont {Goodenough}},
  \bibinfo {author} {\bibfnamefont {H.~D.}\ \bibnamefont {Zhou}}, \bibinfo
  {author} {\bibfnamefont {K.}~\bibnamefont {Matsubayashi}}, \bibinfo {author}
  {\bibfnamefont {Y.}~\bibnamefont {Uwatoko}}, \bibinfo {author} {\bibfnamefont
  {P.~P.}\ \bibnamefont {Kong}}, \bibinfo {author} {\bibfnamefont {C.~Q.}\
  \bibnamefont {Jin}}, \bibinfo {author} {\bibfnamefont {W.~G.}\ \bibnamefont
  {Yang}},\ and\ \bibinfo {author} {\bibfnamefont {G.~Y.}\ \bibnamefont
  {Shen}},\ }\bibfield  {title} {\bibinfo {title} {Pressure effect on the
  structural transition and suppression of the high-spin state in the
  triple-layer
  ${T}^{\ensuremath{'}}\mathrm{\text{\ensuremath{-}}}{\mathrm{la}}_{4}{\mathrm{ni}}_{3}{\mathbf{o}}_{8}$},\
  }\href {https://doi.org/10.1103/PhysRevLett.108.236403} {\bibfield  {journal}
  {\bibinfo  {journal} {Phys. Rev. Lett.}\ }\textbf {\bibinfo {volume} {108}},\
  \bibinfo {pages} {236403} (\bibinfo {year} {2012})}\BibitemShut {NoStop}%
\bibitem [{\citenamefont {Horio}\ \emph {et~al.}(2018)\citenamefont {Horio},
  \citenamefont {Hauser}, \citenamefont {Sassa}, \citenamefont {Mingazheva},
  \citenamefont {Sutter}, \citenamefont {Kramer}, \citenamefont {Cook},
  \citenamefont {Nocerino}, \citenamefont {Forslund}, \citenamefont
  {Tjernberg}, \citenamefont {Kobayashi}, \citenamefont {Chikina},
  \citenamefont {Schröter}, \citenamefont {Krieger}, \citenamefont {Schmitt},
  \citenamefont {Strocov}, \citenamefont {Pyon}, \citenamefont {Takayama},
  \citenamefont {Takagi}, \citenamefont {Lipscombe}, \citenamefont {Hayden},
  \citenamefont {Ishikado}, \citenamefont {Eisaki}, \citenamefont {Neupert},
  \citenamefont {Månsson}, \citenamefont {Matt},\ and\ \citenamefont
  {Chang}}]{Horio2018ThreeDimensional}%
  \BibitemOpen
  \bibfield  {author} {\bibinfo {author} {\bibfnamefont {M.}~\bibnamefont
  {Horio}}, \bibinfo {author} {\bibfnamefont {K.}~\bibnamefont {Hauser}},
  \bibinfo {author} {\bibfnamefont {Y.}~\bibnamefont {Sassa}}, \bibinfo
  {author} {\bibfnamefont {Z.}~\bibnamefont {Mingazheva}}, \bibinfo {author}
  {\bibfnamefont {D.}~\bibnamefont {Sutter}}, \bibinfo {author} {\bibfnamefont
  {K.}~\bibnamefont {Kramer}}, \bibinfo {author} {\bibfnamefont
  {A.}~\bibnamefont {Cook}}, \bibinfo {author} {\bibfnamefont {E.}~\bibnamefont
  {Nocerino}}, \bibinfo {author} {\bibfnamefont {O.~K.}\ \bibnamefont
  {Forslund}}, \bibinfo {author} {\bibfnamefont {O.}~\bibnamefont {Tjernberg}},
  \bibinfo {author} {\bibfnamefont {M.}~\bibnamefont {Kobayashi}}, \bibinfo
  {author} {\bibfnamefont {A.}~\bibnamefont {Chikina}}, \bibinfo {author}
  {\bibfnamefont {N.~B.~M.}\ \bibnamefont {Schröter}}, \bibinfo {author}
  {\bibfnamefont {J.~A.}\ \bibnamefont {Krieger}}, \bibinfo {author}
  {\bibfnamefont {T.}~\bibnamefont {Schmitt}}, \bibinfo {author} {\bibfnamefont
  {V.~N.}\ \bibnamefont {Strocov}}, \bibinfo {author} {\bibfnamefont
  {S.}~\bibnamefont {Pyon}}, \bibinfo {author} {\bibfnamefont {T.}~\bibnamefont
  {Takayama}}, \bibinfo {author} {\bibfnamefont {H.}~\bibnamefont {Takagi}},
  \bibinfo {author} {\bibfnamefont {O.~J.}\ \bibnamefont {Lipscombe}}, \bibinfo
  {author} {\bibfnamefont {S.~M.}\ \bibnamefont {Hayden}}, \bibinfo {author}
  {\bibfnamefont {M.}~\bibnamefont {Ishikado}}, \bibinfo {author}
  {\bibfnamefont {H.}~\bibnamefont {Eisaki}}, \bibinfo {author} {\bibfnamefont
  {T.}~\bibnamefont {Neupert}}, \bibinfo {author} {\bibfnamefont
  {M.}~\bibnamefont {Månsson}}, \bibinfo {author} {\bibfnamefont {C.~E.}\
  \bibnamefont {Matt}},\ and\ \bibinfo {author} {\bibfnamefont
  {J.}~\bibnamefont {Chang}},\ }\bibfield  {title} {\bibinfo {title}
  {Three-dimensional fermi surface of overdoped la-based cuprates},\ }\href
  {https://doi.org/10.1103/PhysRevLett.121.077004} {\bibfield  {journal}
  {\bibinfo  {journal} {Physical Review Letters}\ }\textbf {\bibinfo {volume}
  {121}},\ \bibinfo {pages} {077004} (\bibinfo {year} {2018})}\BibitemShut
  {NoStop}%
\bibitem [{\citenamefont {Daou}\ \emph {et~al.}(2009)\citenamefont {Daou},
  \citenamefont {Doiron-Leyraud}, \citenamefont {LeBoeuf}, \citenamefont {Li},
  \citenamefont {Lalibert\'{e}}, \citenamefont {Cyr-Choini\`{e}re},
  \citenamefont {Jo}, \citenamefont {Balicas}, \citenamefont {Yan},
  \citenamefont {Zhou}, \citenamefont {Goodenough},\ and\ \citenamefont
  {Taillefer}}]{Daou2009Linear}%
  \BibitemOpen
  \bibfield  {author} {\bibinfo {author} {\bibfnamefont {R.}~\bibnamefont
  {Daou}}, \bibinfo {author} {\bibfnamefont {N.}~\bibnamefont
  {Doiron-Leyraud}}, \bibinfo {author} {\bibfnamefont {D.}~\bibnamefont
  {LeBoeuf}}, \bibinfo {author} {\bibfnamefont {S.~Y.}\ \bibnamefont {Li}},
  \bibinfo {author} {\bibfnamefont {F.}~\bibnamefont {Lalibert\'{e}}}, \bibinfo
  {author} {\bibfnamefont {O.}~\bibnamefont {Cyr-Choini\`{e}re}}, \bibinfo
  {author} {\bibfnamefont {Y.~J.}\ \bibnamefont {Jo}}, \bibinfo {author}
  {\bibfnamefont {L.}~\bibnamefont {Balicas}}, \bibinfo {author} {\bibfnamefont
  {J.-Q.}\ \bibnamefont {Yan}}, \bibinfo {author} {\bibfnamefont {J.-S.}\
  \bibnamefont {Zhou}}, \bibinfo {author} {\bibfnamefont {J.~B.}\ \bibnamefont
  {Goodenough}},\ and\ \bibinfo {author} {\bibfnamefont {L.}~\bibnamefont
  {Taillefer}},\ }\bibfield  {title} {\bibinfo {title} {Linear temperature
  dependence of resistivity and change in the fermi surface at the pseudogap
  critical point of a high-${T}_c$ superconductor},\ }\href
  {https://doi.org/10.1038/nphys1109} {\bibfield  {journal} {\bibinfo
  {journal} {Nature Physics}\ }\textbf {\bibinfo {volume} {5}},\ \bibinfo
  {pages} {31} (\bibinfo {year} {2009})}\BibitemShut {NoStop}%
\bibitem [{\citenamefont {Kondo}\ \emph {et~al.}(2006)\citenamefont {Kondo},
  \citenamefont {Takeuchi}, \citenamefont {Tsuda},\ and\ \citenamefont
  {Shin}}]{kondo_resistivity2006}%
  \BibitemOpen
  \bibfield  {author} {\bibinfo {author} {\bibfnamefont {T.}~\bibnamefont
  {Kondo}}, \bibinfo {author} {\bibfnamefont {T.}~\bibnamefont {Takeuchi}},
  \bibinfo {author} {\bibfnamefont {S.}~\bibnamefont {Tsuda}},\ and\ \bibinfo
  {author} {\bibfnamefont {S.}~\bibnamefont {Shin}},\ }\bibfield  {title}
  {\bibinfo {title} {Electrical resistivity and scattering processes in
  ${(\mathrm{Bi},\mathrm{Pb})}_{2}{(\mathrm{Sr},\mathrm{La})}_{2}\mathrm{Cu}{\mathrm{o}}_{6+\ensuremath{\delta}}$
  studied by angle-resolved photoemission spectroscopy},\ }\href
  {https://doi.org/10.1103/PhysRevB.74.224511} {\bibfield  {journal} {\bibinfo
  {journal} {Phys. Rev. B}\ }\textbf {\bibinfo {volume} {74}},\ \bibinfo
  {pages} {224511} (\bibinfo {year} {2006})}\BibitemShut {NoStop}%
\bibitem [{\citenamefont {Ayres}\ \emph {et~al.}(2021)\citenamefont {Ayres},
  \citenamefont {Berben}, \citenamefont {Čulo}, \citenamefont {Hsu},
  \citenamefont {van Heumen}, \citenamefont {Huang}, \citenamefont {Zaanen},
  \citenamefont {Kondo}, \citenamefont {Takeuchi}, \citenamefont {Cooper},
  \citenamefont {Putzke}, \citenamefont {Friedemann}, \citenamefont
  {Carrington},\ and\ \citenamefont {Hussey}}]{Ayres2021Incoherent}%
  \BibitemOpen
  \bibfield  {author} {\bibinfo {author} {\bibfnamefont {J.}~\bibnamefont
  {Ayres}}, \bibinfo {author} {\bibfnamefont {M.}~\bibnamefont {Berben}},
  \bibinfo {author} {\bibfnamefont {M.}~\bibnamefont {Čulo}}, \bibinfo
  {author} {\bibfnamefont {Y.-T.}\ \bibnamefont {Hsu}}, \bibinfo {author}
  {\bibfnamefont {E.}~\bibnamefont {van Heumen}}, \bibinfo {author}
  {\bibfnamefont {Y.}~\bibnamefont {Huang}}, \bibinfo {author} {\bibfnamefont
  {J.}~\bibnamefont {Zaanen}}, \bibinfo {author} {\bibfnamefont
  {T.}~\bibnamefont {Kondo}}, \bibinfo {author} {\bibfnamefont
  {T.}~\bibnamefont {Takeuchi}}, \bibinfo {author} {\bibfnamefont {J.~R.}\
  \bibnamefont {Cooper}}, \bibinfo {author} {\bibfnamefont {C.}~\bibnamefont
  {Putzke}}, \bibinfo {author} {\bibfnamefont {S.}~\bibnamefont {Friedemann}},
  \bibinfo {author} {\bibfnamefont {A.}~\bibnamefont {Carrington}},\ and\
  \bibinfo {author} {\bibfnamefont {N.~E.}\ \bibnamefont {Hussey}},\ }\bibfield
   {title} {\bibinfo {title} {Incoherent transport across the strange-metal
  regime of overdoped cuprates},\ }\href
  {https://doi.org/10.1038/s41586-021-03622-z} {\bibfield  {journal} {\bibinfo
  {journal} {Nature}\ }\textbf {\bibinfo {volume} {595}},\ \bibinfo {pages}
  {661} (\bibinfo {year} {2021})}\BibitemShut {NoStop}%
\bibitem [{\citenamefont {Jin}\ \emph {et~al.}(2021)\citenamefont {Jin},
  \citenamefont {Narduzzo}, \citenamefont {Nohara}, \citenamefont {Takagi},
  \citenamefont {Hussey},\ and\ \citenamefont {Behnia}}]{jin2021}%
  \BibitemOpen
  \bibfield  {author} {\bibinfo {author} {\bibfnamefont {H.}~\bibnamefont
  {Jin}}, \bibinfo {author} {\bibfnamefont {A.}~\bibnamefont {Narduzzo}},
  \bibinfo {author} {\bibfnamefont {M.}~\bibnamefont {Nohara}}, \bibinfo
  {author} {\bibfnamefont {H.}~\bibnamefont {Takagi}}, \bibinfo {author}
  {\bibfnamefont {N.~E.}\ \bibnamefont {Hussey}},\ and\ \bibinfo {author}
  {\bibfnamefont {K.}~\bibnamefont {Behnia}},\ }\bibfield  {title} {\bibinfo
  {title} {Positive {Seebeck} coefficient in highly doped
  {La}$_{2-x}${Sr}$_x${CuO}$_4$ ($x$=0.33); its origin and implication},\
  }\href {https://doi.org/10.7566/JPSJ.90.053702} {\bibfield  {journal}
  {\bibinfo  {journal} {Journal of the Physical Society of Japan}\ }\textbf
  {\bibinfo {volume} {90}},\ \bibinfo {pages} {053702} (\bibinfo {year}
  {2021})}\BibitemShut {NoStop}%
\bibitem [{\citenamefont {Quirk}\ \emph {et~al.}(2023)\citenamefont {Quirk},
  \citenamefont {Li}, \citenamefont {Wang}, \citenamefont {Hwang},\ and\
  \citenamefont {Ong}}]{quirk2023}%
  \BibitemOpen
  \bibfield  {author} {\bibinfo {author} {\bibfnamefont {N.~P.}\ \bibnamefont
  {Quirk}}, \bibinfo {author} {\bibfnamefont {D.}~\bibnamefont {Li}}, \bibinfo
  {author} {\bibfnamefont {B.~Y.}\ \bibnamefont {Wang}}, \bibinfo {author}
  {\bibfnamefont {H.~Y.}\ \bibnamefont {Hwang}},\ and\ \bibinfo {author}
  {\bibfnamefont {N.~P.}\ \bibnamefont {Ong}},\ }\href
  {https://arxiv.org/abs/2309.03170} {\bibinfo {title} {The vortex-nernst
  effect in a superconducting infinite-layer nickelate}} (\bibinfo {year}
  {2023}),\ \Eprint {https://arxiv.org/abs/2309.03170} {arXiv:2309.03170
  [cond-mat.supr-con]} \BibitemShut {NoStop}%
\bibitem [{\citenamefont {Yordanov}\ \emph {et~al.}(2019)\citenamefont
  {Yordanov}, \citenamefont {Sigle}, \citenamefont {Kaya}, \citenamefont
  {Gruner}, \citenamefont {Pentcheva}, \citenamefont {Keimer},\ and\
  \citenamefont {Habermeier}}]{Yordanov2019Large}%
  \BibitemOpen
  \bibfield  {author} {\bibinfo {author} {\bibfnamefont {P.}~\bibnamefont
  {Yordanov}}, \bibinfo {author} {\bibfnamefont {W.}~\bibnamefont {Sigle}},
  \bibinfo {author} {\bibfnamefont {P.}~\bibnamefont {Kaya}}, \bibinfo {author}
  {\bibfnamefont {M.~E.}\ \bibnamefont {Gruner}}, \bibinfo {author}
  {\bibfnamefont {R.}~\bibnamefont {Pentcheva}}, \bibinfo {author}
  {\bibfnamefont {B.}~\bibnamefont {Keimer}},\ and\ \bibinfo {author}
  {\bibfnamefont {H.-U.}\ \bibnamefont {Habermeier}},\ }\bibfield  {title}
  {\bibinfo {title} {Large thermopower anisotropy in pdco o 2 thin films},\
  }\href {https://doi.org/10.1103/PhysRevMaterials.3.085403} {\bibfield
  {journal} {\bibinfo  {journal} {Physical Review Materials}\ }\textbf
  {\bibinfo {volume} {3}},\ \bibinfo {pages} {085403} (\bibinfo {year}
  {2019})}\BibitemShut {NoStop}%
\bibitem [{\citenamefont {Sun}\ \emph {et~al.}(2024)\citenamefont {Sun},
  \citenamefont {Jiang}, \citenamefont {Xia}, \citenamefont {Hao},
  \citenamefont {Li}, \citenamefont {Yan}, \citenamefont {Wang}, \citenamefont
  {Liu}, \citenamefont {Ding}, \citenamefont {Liu}, \citenamefont {Liu},
  \citenamefont {Liu}, \citenamefont {Chen}, \citenamefont {Shen},\ and\
  \citenamefont {Nie}}]{sun_2024}%
  \BibitemOpen
  \bibfield  {author} {\bibinfo {author} {\bibfnamefont {W.}~\bibnamefont
  {Sun}}, \bibinfo {author} {\bibfnamefont {Z.}~\bibnamefont {Jiang}}, \bibinfo
  {author} {\bibfnamefont {C.}~\bibnamefont {Xia}}, \bibinfo {author}
  {\bibfnamefont {B.}~\bibnamefont {Hao}}, \bibinfo {author} {\bibfnamefont
  {Y.}~\bibnamefont {Li}}, \bibinfo {author} {\bibfnamefont {S.}~\bibnamefont
  {Yan}}, \bibinfo {author} {\bibfnamefont {M.}~\bibnamefont {Wang}}, \bibinfo
  {author} {\bibfnamefont {H.}~\bibnamefont {Liu}}, \bibinfo {author}
  {\bibfnamefont {J.}~\bibnamefont {Ding}}, \bibinfo {author} {\bibfnamefont
  {J.}~\bibnamefont {Liu}}, \bibinfo {author} {\bibfnamefont {Z.}~\bibnamefont
  {Liu}}, \bibinfo {author} {\bibfnamefont {J.}~\bibnamefont {Liu}}, \bibinfo
  {author} {\bibfnamefont {H.}~\bibnamefont {Chen}}, \bibinfo {author}
  {\bibfnamefont {D.}~\bibnamefont {Shen}},\ and\ \bibinfo {author}
  {\bibfnamefont {Y.}~\bibnamefont {Nie}},\ }\href
  {https://doi.org/10.48550/arXiv.2403.07344} {\bibinfo {title} {Electronic
  {{Structure}} of {{Superconducting Infinite-Layer Lanthanum Nickelates}}}}
  (\bibinfo {year} {2024}),\ \Eprint {https://arxiv.org/abs/2403.07344}
  {2403.07344 [cond-mat]} \BibitemShut {NoStop}%
\bibitem [{\citenamefont {Sun}\ \emph {et~al.}(2023)\citenamefont {Sun},
  \citenamefont {Huo}, \citenamefont {Hu}, \citenamefont {Li}, \citenamefont
  {Liu}, \citenamefont {Han}, \citenamefont {Tang}, \citenamefont {Mao},
  \citenamefont {Yang}, \citenamefont {Wang}, \citenamefont {Cheng},
  \citenamefont {Yao}, \citenamefont {Zhang},\ and\ \citenamefont
  {Wang}}]{sun_2023}%
  \BibitemOpen
  \bibfield  {author} {\bibinfo {author} {\bibfnamefont {H.}~\bibnamefont
  {Sun}}, \bibinfo {author} {\bibfnamefont {M.}~\bibnamefont {Huo}}, \bibinfo
  {author} {\bibfnamefont {X.}~\bibnamefont {Hu}}, \bibinfo {author}
  {\bibfnamefont {J.}~\bibnamefont {Li}}, \bibinfo {author} {\bibfnamefont
  {Z.}~\bibnamefont {Liu}}, \bibinfo {author} {\bibfnamefont {Y.}~\bibnamefont
  {Han}}, \bibinfo {author} {\bibfnamefont {L.}~\bibnamefont {Tang}}, \bibinfo
  {author} {\bibfnamefont {Z.}~\bibnamefont {Mao}}, \bibinfo {author}
  {\bibfnamefont {P.}~\bibnamefont {Yang}}, \bibinfo {author} {\bibfnamefont
  {B.}~\bibnamefont {Wang}}, \bibinfo {author} {\bibfnamefont {J.}~\bibnamefont
  {Cheng}}, \bibinfo {author} {\bibfnamefont {D.-X.}\ \bibnamefont {Yao}},
  \bibinfo {author} {\bibfnamefont {G.-M.}\ \bibnamefont {Zhang}},\ and\
  \bibinfo {author} {\bibfnamefont {M.}~\bibnamefont {Wang}},\ }\bibfield
  {title} {\bibinfo {title} {Signatures of superconductivity near 80 {{K}} in a
  nickelate under high pressure},\ }\href
  {https://doi.org/10.1038/s41586-023-06408-7} {\bibfield  {journal} {\bibinfo
  {journal} {Nature}\ }\textbf {\bibinfo {volume} {621}},\ \bibinfo {pages}
  {493} (\bibinfo {year} {2023})}\BibitemShut {NoStop}%
\bibitem [{\citenamefont {Hartnoll}\ and\ \citenamefont
  {Mackenzie}(2022)}]{hartnoll2021planckian}%
  \BibitemOpen
  \bibfield  {author} {\bibinfo {author} {\bibfnamefont {S.~A.}\ \bibnamefont
  {Hartnoll}}\ and\ \bibinfo {author} {\bibfnamefont {A.~P.}\ \bibnamefont
  {Mackenzie}},\ }\bibfield  {title} {\bibinfo {title} {Colloquium: Planckian
  dissipation in metals},\ }\href
  {https://doi.org/10.1103/RevModPhys.94.041002} {\bibfield  {journal}
  {\bibinfo  {journal} {Rev. Mod. Phys.}\ }\textbf {\bibinfo {volume} {94}},\
  \bibinfo {pages} {041002} (\bibinfo {year} {2022})}\BibitemShut {NoStop}%
\bibitem [{\citenamefont {Perdew}\ \emph {et~al.}(1996)\citenamefont {Perdew},
  \citenamefont {Burke},\ and\ \citenamefont
  {Ernzerhof}}]{Perdew1996generalized}%
  \BibitemOpen
  \bibfield  {author} {\bibinfo {author} {\bibfnamefont {J.~P.}\ \bibnamefont
  {Perdew}}, \bibinfo {author} {\bibfnamefont {K.}~\bibnamefont {Burke}},\ and\
  \bibinfo {author} {\bibfnamefont {M.}~\bibnamefont {Ernzerhof}},\ }\bibfield
  {title} {\bibinfo {title} {Generalized gradient approximation made simple},\
  }\href {https://doi.org/10.1103/PhysRevLett.77.3865} {\bibfield  {journal}
  {\bibinfo  {journal} {Phys. Rev. Lett.}\ }\textbf {\bibinfo {volume} {77}},\
  \bibinfo {pages} {3865} (\bibinfo {year} {1996})}\BibitemShut {NoStop}%
\bibitem [{\citenamefont {Markiewicz}\ \emph {et~al.}(2005)\citenamefont
  {Markiewicz}, \citenamefont {Sahrakorpi}, \citenamefont {Lindroos},
  \citenamefont {Lin},\ and\ \citenamefont {Bansil}}]{Markiewicz2005OneBand}%
  \BibitemOpen
  \bibfield  {author} {\bibinfo {author} {\bibfnamefont {R.~S.}\ \bibnamefont
  {Markiewicz}}, \bibinfo {author} {\bibfnamefont {S.}~\bibnamefont
  {Sahrakorpi}}, \bibinfo {author} {\bibfnamefont {M.}~\bibnamefont
  {Lindroos}}, \bibinfo {author} {\bibfnamefont {H.}~\bibnamefont {Lin}},\ and\
  \bibinfo {author} {\bibfnamefont {A.}~\bibnamefont {Bansil}},\ }\bibfield
  {title} {\bibinfo {title} {One-band tight-binding model parametrization of
  the high-${T}_c$ cuprates including the effect of $k_z$ dispersion},\ }\href
  {https://doi.org/10.1103/PhysRevB.72.054519} {\bibfield  {journal} {\bibinfo
  {journal} {Physical Review B}\ }\textbf {\bibinfo {volume} {72}},\ \bibinfo
  {pages} {054519} (\bibinfo {year} {2005})}\BibitemShut {NoStop}%
\bibitem [{\citenamefont {Li}\ and\ \citenamefont {Louie}(2022)}]{Louie2022}%
  \BibitemOpen
  \bibfield  {author} {\bibinfo {author} {\bibfnamefont {Z.}~\bibnamefont
  {Li}}\ and\ \bibinfo {author} {\bibfnamefont {S.~G.}\ \bibnamefont {Louie}},\
  }\bibfield  {title} {\bibinfo {title} {Two-gap superconductivity and decisive
  role of rare-earth $d$ electrons in infinite-layer nickelates},\ }\href
  {https://arxiv.org/abs/2210.12819} {\bibfield  {journal} {\bibinfo  {journal}
  {arXiv:2210.12819}\ } (\bibinfo {year} {2022})}\BibitemShut {NoStop}%
\bibitem [{\citenamefont {Bruin}\ \emph {et~al.}(2013)\citenamefont {Bruin},
  \citenamefont {Sakai}, \citenamefont {Perry},\ and\ \citenamefont
  {Mackenzie}}]{Bruin2013Similarity}%
  \BibitemOpen
  \bibfield  {author} {\bibinfo {author} {\bibfnamefont {J.~a.~N.}\
  \bibnamefont {Bruin}}, \bibinfo {author} {\bibfnamefont {H.}~\bibnamefont
  {Sakai}}, \bibinfo {author} {\bibfnamefont {R.~S.}\ \bibnamefont {Perry}},\
  and\ \bibinfo {author} {\bibfnamefont {A.~P.}\ \bibnamefont {Mackenzie}},\
  }\bibfield  {title} {\bibinfo {title} {Similarity of scattering rates in
  metals showing t-linear resistivity},\ }\href
  {https://doi.org/10.1126/science.1227612} {\bibfield  {journal} {\bibinfo
  {journal} {Science}\ }\textbf {\bibinfo {volume} {339}},\ \bibinfo {pages}
  {804} (\bibinfo {year} {2013})}\BibitemShut {NoStop}%
\bibitem [{\citenamefont {Legros}\ \emph {et~al.}(2019)\citenamefont {Legros},
  \citenamefont {Benhabib}, \citenamefont {Tabis}, \citenamefont {Laliberté},
  \citenamefont {Dion}, \citenamefont {Lizaire}, \citenamefont {Vignolle},
  \citenamefont {Vignolles}, \citenamefont {Raffy}, \citenamefont {Li},
  \citenamefont {Auban-Senzier}, \citenamefont {Doiron-Leyraud}, \citenamefont
  {Fournier}, \citenamefont {Colson}, \citenamefont {Taillefer},\ and\
  \citenamefont {Proust}}]{Legros2019Universal}%
  \BibitemOpen
  \bibfield  {author} {\bibinfo {author} {\bibfnamefont {A.}~\bibnamefont
  {Legros}}, \bibinfo {author} {\bibfnamefont {S.}~\bibnamefont {Benhabib}},
  \bibinfo {author} {\bibfnamefont {W.}~\bibnamefont {Tabis}}, \bibinfo
  {author} {\bibfnamefont {F.}~\bibnamefont {Laliberté}}, \bibinfo {author}
  {\bibfnamefont {M.}~\bibnamefont {Dion}}, \bibinfo {author} {\bibfnamefont
  {M.}~\bibnamefont {Lizaire}}, \bibinfo {author} {\bibfnamefont
  {B.}~\bibnamefont {Vignolle}}, \bibinfo {author} {\bibfnamefont
  {D.}~\bibnamefont {Vignolles}}, \bibinfo {author} {\bibfnamefont
  {H.}~\bibnamefont {Raffy}}, \bibinfo {author} {\bibfnamefont {Z.~Z.}\
  \bibnamefont {Li}}, \bibinfo {author} {\bibfnamefont {P.}~\bibnamefont
  {Auban-Senzier}}, \bibinfo {author} {\bibfnamefont {N.}~\bibnamefont
  {Doiron-Leyraud}}, \bibinfo {author} {\bibfnamefont {P.}~\bibnamefont
  {Fournier}}, \bibinfo {author} {\bibfnamefont {D.}~\bibnamefont {Colson}},
  \bibinfo {author} {\bibfnamefont {L.}~\bibnamefont {Taillefer}},\ and\
  \bibinfo {author} {\bibfnamefont {C.}~\bibnamefont {Proust}},\ }\bibfield
  {title} {\bibinfo {title} {Universal t -linear resistivity and planckian
  dissipation in overdoped cuprates},\ }\href
  {https://doi.org/10.1038/s41567-018-0334-2} {\bibfield  {journal} {\bibinfo
  {journal} {Nature Physics}\ }\textbf {\bibinfo {volume} {15}},\ \bibinfo
  {pages} {142} (\bibinfo {year} {2019})}\BibitemShut {NoStop}%
\bibitem [{\citenamefont {Fang}\ \emph {et~al.}(2009)\citenamefont {Fang},
  \citenamefont {Luo}, \citenamefont {Cheng}, \citenamefont {Wang},
  \citenamefont {Jia}, \citenamefont {Mu}, \citenamefont {Shen}, \citenamefont
  {Mazin}, \citenamefont {Shan}, \citenamefont {Ren},\ and\ \citenamefont
  {Wen}}]{fang2009}%
  \BibitemOpen
  \bibfield  {author} {\bibinfo {author} {\bibfnamefont {L.}~\bibnamefont
  {Fang}}, \bibinfo {author} {\bibfnamefont {H.}~\bibnamefont {Luo}}, \bibinfo
  {author} {\bibfnamefont {P.}~\bibnamefont {Cheng}}, \bibinfo {author}
  {\bibfnamefont {Z.}~\bibnamefont {Wang}}, \bibinfo {author} {\bibfnamefont
  {Y.}~\bibnamefont {Jia}}, \bibinfo {author} {\bibfnamefont {G.}~\bibnamefont
  {Mu}}, \bibinfo {author} {\bibfnamefont {B.}~\bibnamefont {Shen}}, \bibinfo
  {author} {\bibfnamefont {I.~I.}\ \bibnamefont {Mazin}}, \bibinfo {author}
  {\bibfnamefont {L.}~\bibnamefont {Shan}}, \bibinfo {author} {\bibfnamefont
  {C.}~\bibnamefont {Ren}},\ and\ \bibinfo {author} {\bibfnamefont {H.-H.}\
  \bibnamefont {Wen}},\ }\bibfield  {title} {\bibinfo {title} {Roles of
  multiband effects and electron-hole asymmetry in the superconductivity and
  normal-state properties of
  $\text{Ba}{({\text{Fe}}_{1\ensuremath{-}x}{\text{Co}}_{x})}_{2}{\text{as}}_{2}$},\
  }\href {https://doi.org/10.1103/PhysRevB.80.140508} {\bibfield  {journal}
  {\bibinfo  {journal} {Phys. Rev. B}\ }\textbf {\bibinfo {volume} {80}},\
  \bibinfo {pages} {140508} (\bibinfo {year} {2009})}\BibitemShut {NoStop}%
\bibitem [{\citenamefont {Doiron-Leyraud}\ \emph {et~al.}(2009)\citenamefont
  {Doiron-Leyraud}, \citenamefont {Auban-Senzier}, \citenamefont {Ren\'e~de
  Cotret}, \citenamefont {Bourbonnais}, \citenamefont {J\'erome}, \citenamefont
  {Bechgaard},\ and\ \citenamefont {Taillefer}}]{doiron-leyraud2009}%
  \BibitemOpen
  \bibfield  {author} {\bibinfo {author} {\bibfnamefont {N.}~\bibnamefont
  {Doiron-Leyraud}}, \bibinfo {author} {\bibfnamefont {P.}~\bibnamefont
  {Auban-Senzier}}, \bibinfo {author} {\bibfnamefont {S.}~\bibnamefont
  {Ren\'e~de Cotret}}, \bibinfo {author} {\bibfnamefont {C.}~\bibnamefont
  {Bourbonnais}}, \bibinfo {author} {\bibfnamefont {D.}~\bibnamefont
  {J\'erome}}, \bibinfo {author} {\bibfnamefont {K.}~\bibnamefont
  {Bechgaard}},\ and\ \bibinfo {author} {\bibfnamefont {L.}~\bibnamefont
  {Taillefer}},\ }\bibfield  {title} {\bibinfo {title} {Correlation between
  linear resistivity and ${T}_{c}$ in the bechgaard salts and the pnictide
  superconductor
  $\text{Ba}{({\text{Fe}}_{1\ensuremath{-}x}{\text{Co}}_{x})}_{2}{\text{as}}_{2}$},\
  }\href {https://doi.org/10.1103/PhysRevB.80.214531} {\bibfield  {journal}
  {\bibinfo  {journal} {Phys. Rev. B}\ }\textbf {\bibinfo {volume} {80}},\
  \bibinfo {pages} {214531} (\bibinfo {year} {2009})}\BibitemShut {NoStop}%
\bibitem [{\citenamefont {Badoux}\ \emph {et~al.}(2016)\citenamefont {Badoux},
  \citenamefont {Afshar}, \citenamefont {Michon}, \citenamefont {Ouellet},
  \citenamefont {Fortier}, \citenamefont {LeBoeuf}, \citenamefont {Croft},
  \citenamefont {Lester}, \citenamefont {Hayden}, \citenamefont {Takagi},
  \citenamefont {Yamada}, \citenamefont {Graf}, \citenamefont
  {Doiron-Leyraud},\ and\ \citenamefont {Taillefer}}]{badoux2016}%
  \BibitemOpen
  \bibfield  {author} {\bibinfo {author} {\bibfnamefont {S.}~\bibnamefont
  {Badoux}}, \bibinfo {author} {\bibfnamefont {S.~A.~A.}\ \bibnamefont
  {Afshar}}, \bibinfo {author} {\bibfnamefont {B.}~\bibnamefont {Michon}},
  \bibinfo {author} {\bibfnamefont {A.}~\bibnamefont {Ouellet}}, \bibinfo
  {author} {\bibfnamefont {S.}~\bibnamefont {Fortier}}, \bibinfo {author}
  {\bibfnamefont {D.}~\bibnamefont {LeBoeuf}}, \bibinfo {author} {\bibfnamefont
  {T.~P.}\ \bibnamefont {Croft}}, \bibinfo {author} {\bibfnamefont
  {C.}~\bibnamefont {Lester}}, \bibinfo {author} {\bibfnamefont {S.~M.}\
  \bibnamefont {Hayden}}, \bibinfo {author} {\bibfnamefont {H.}~\bibnamefont
  {Takagi}}, \bibinfo {author} {\bibfnamefont {K.}~\bibnamefont {Yamada}},
  \bibinfo {author} {\bibfnamefont {D.}~\bibnamefont {Graf}}, \bibinfo {author}
  {\bibfnamefont {N.}~\bibnamefont {Doiron-Leyraud}},\ and\ \bibinfo {author}
  {\bibfnamefont {L.}~\bibnamefont {Taillefer}},\ }\bibfield  {title} {\bibinfo
  {title} {Critical doping for the onset of fermi-surface reconstruction by
  charge-density-wave order in the cuprate superconductor
  ${\mathrm{la}}_{2\ensuremath{-}x}{\mathrm{sr}}_{x}{\mathrm{cuo}}_{4}$},\
  }\href {https://doi.org/10.1103/PhysRevX.6.021004} {\bibfield  {journal}
  {\bibinfo  {journal} {Phys. Rev. X}\ }\textbf {\bibinfo {volume} {6}},\
  \bibinfo {pages} {021004} (\bibinfo {year} {2016})}\BibitemShut {NoStop}%
\end{thebibliography}


\end{document}